\documentclass[11pt,a4paper]{article}
\usepackage{jheppub}
\pdfoutput=1

\usepackage[]{amsmath}
\usepackage[]{amssymb}
\usepackage{amsfonts}

\DeclareMathOperator{\Z}{\mathbb{Z}}
\DeclareMathOperator{\cs}{\mathbb{S}}

\newcommand{\I}{\mathrm{i}}
\newcommand{\dd}{\mathrm{d}}

\newcommand{\fre}{\mathcal{F}}
\newcommand{\fee}{\widetilde{\mathcal{F}}}
\newcommand{\oR}{\mathcal{O} \left( 1 / R \right)}

\newcommand{\cp}{C_{\mathrm{P}}}
\newcommand{\cI}{C_{\mathrm{I}}}
\newcommand{\cII}{C_{\mathrm{II}}}
\newcommand{\cIII}{C_{\mathrm{III}}}
\newcommand{\cIV}{C_{\mathrm{IV}}}
\newcommand{\cV}{C_{\mathrm{V}}}

\newcommand{\re}[1]{\mathrm{Re} \left( #1 \right)}
\newcommand{\im}[1]{\mathrm{Im} \left( #1 \right)}

\def\Xint#1{\mathchoice
   {\XXint\displaystyle\textstyle{#1}}%
   {\XXint\textstyle\scriptstyle{#1}}%
   {\XXint\scriptstyle\scriptscriptstyle{#1}}%
   {\XXint\scriptscriptstyle\scriptscriptstyle{#1}}%
   \!\int}
\def\XXint#1#2#3{{\setbox0=\hbox{$#1{#2#3}{\int}$}
     \vcenter{\hbox{$#2#3$}}\kern-.52\wd0}}
\def\dashint{\Xint-}

\usepackage{hyperref}

\usepackage[toc,page]{appendix}

\usepackage[utf8]{inputenc}
\usepackage[english]{babel}

 \usepackage{graphicx}
 
 \usepackage{enumerate}

\title{Phase transitions and Wilson loops in antisymmetric representations in Chern--Simons-matter theory}
\author{Leonardo Santilli}
\author{and Miguel Tierz}
\affiliation{Departamento de Matem\'{a}tica, Grupo de F\'{\i}sica Matem\'{a}tica, Faculdade de Ci\^{e}ncias, Universidade de Lisboa, Campo Grande, Edif\'{\i}cio C6, 1749-016 Lisboa, Portugal.}
\emailAdd{lsantilli@fc.ul.pt}
\emailAdd{tierz@fc.ul.pt}

\abstract{We study the phase transitions of three-dimensional $\mathcal{N}=2$ $U(N)$ Chern--Simons theory on $\mathbb{S}^3$ with a varied number of massive fundamental hypermultiplets and with a Fayet--Iliopoulos parameter. We characterize the various phase diagrams in the decompactification limit, according to the number of different mass scales in the theory. For this, we extend the known solution of the saddle-point equations to the setting where the one cut solution is characterized by asymmetric intervals. We then study the large $N$ limit of Wilson loops in antisymmetric representations, with the additional scaling corresponding to the variation of the size of the representation. We give explicit expressions, both with and without FI terms, and study $1/R$ corrections for the different phases. These corrections break the perimeter law behavior, as they introduce a scaling with the size of the representation. We show how the phase transitions of the Wilson loops can either be of first or second order and determine the underlying mechanism in terms of the eventual asymmetry of the support of the solution of the saddle-point equation, at the critical points.}

\begin{document}

	\maketitle

	\section{Introduction}
	
	The study of supersymmetric gauge theories in compact manifolds has experienced great progress on the last decade, due to the development of the localization method \cite{Pestun:2007rz}, which leads to a much simplified description of the original functional integral describing the observables of the gauge theories. The resulting object to analyze is a finite-dimensional integral of random matrix type. 
	
		An important stream of research, that emerged by analyzing such matrix model description, brought upon by localization, is the discovery and characterization of large $N$ quantum phase transitions of supersymmetric gauge theories, mostly on spheres, of different dimensions, and typically in the decompactification limit of the sphere, to be defined below. Aspects of this research line have been developed by Russo, Zarembo and collaborators, especially in the four-dimensional case, where it was found to be quite a generic feature of $\mathcal{N}=2$ theories with massive matter \cite{Russo:2012ay,Russo:2013qaa,Russo:2013kea,Russo:2013sba,Buchel:2013id,Russo:2014nka,Russo:2014bda,Zarembo:2014ooa,Chen:2014vka}, although there are some exception such as $\mathcal{N}=2$ $SU(N)$ theory with $N_{f}=2N$, a superconformal theory in four dimensions \cite{Russo:2012ay}. 
		
		In these works, it can be seen that the  quantum  critical  points originate from resonances emerging when the coupling is such that the saddle point hits points in the Coulomb branch of the moduli space where there are massless excitations. In other words, there are critical values for the coupling such that, when crossed, field configurations with extra massless multiplets contribute to the saddle-point, leading to discontinuities in vacuum expectation values of supersymmetric observables, such as the partition function and Wilson loops. 
		
		The phase transitions in four dimensions have also been analyzed by studying the holographic dual (see for example \cite{Bobev:2013cja} and \cite{Buchel:2013id} encompassing both approaches), but we will be focusing on the localized matrix model approach and, in addition, we shall study the three-dimensional case \cite{Kapustin:2009kz}. In three dimensions, $U(N)$ Chern--Simons-matter with $2N_{f}$ massive hypermultiplets in the fundamental representation has been studied \cite{Barranco:2014,Russo:2014bda} and also ABJM theory, with mass-deformations \cite{Anderson:2014hxa} (see also \cite{Nosaka:2015bhf,Nosaka:2016vqf}) and with mass-deformation and analytically continued Fayet--Iliopoulos (FI) terms, have been studied \cite{Anderson:2015}, leading to a very rich phase structure. See \cite{Nosaka:2015bhf,Nosaka:2016vqf} for further work on the mass-deformed ABJM theory, and \cite{Honda:2018pqa} for the study of phase transitions from a different perspective.
		
		It is worth mentioning that in three dimensions, phase transitions have been studied, even in the case of Abelian gauge group and without Chern--Simons term, that is for supersymmetric QED on $\cs^3$ \cite{Russo:2016ueu} (see also the different work on sQED, with implications in condensed-matter physics \cite{Jian:2016zll}). See also \cite{Anderson:2017xrv} where the phase structure of a mass-deformed Chern--Simons-matter theory with real masses uniformly distributed along a segment has been obtained. Results in higher dimensions have been obtained in \cite{Nedelin:2015mta,Minahan:2014hwa,Marmiroli:2014ssa}.
		
		We will follow and study first, in exactly the same manner as \cite{Barranco:2014}, the phase structure of $U(N)$ Chern--Simons-matter with massive fundamental hypermultiplets, but for a more general mass configurations and with the presence of a FI parameter as well. Then, we analyze the Wilson loop in the antisymmetric representation at large $N$ in the case of $\mathcal{N}=2$ supersymmetric $U(N)$ Chern--Simons theory, first with the insertion of massive matter \cite{Barranco:2014} and then also introducing a FI deformation term. We adopt the now standard and widespread technique described in \cite{Hartnoll:2006} to find out the explicit form of the Wilson loop. This approach is based on studying, for large $N$ and using the saddle-point method, the matrix model average of the generating function of the Wilson loops in symmetric and antisymmetric representations. In mathematical terms, these are the generating functions of elementary and homogeneous symmetric polynomials. 
		
		The study of Wilson loops in large representations is comparatively less developed than the case of the Wilson loop in the fundamental representation, although in four dimensions, there are a number of works in the higher-rank case in the last years \cite{Russo:2017ngf,Chen-Lin:2015dfa}. The case of a three dimensional Chern--Simons theory is understudied in comparison and we tackle it here. We have now then an additional scaling parameter to play with in the large $N$ and decompactification limits, namely $f=k/N$, where $k$ denotes the size of the partition indexing the Wilson loop representation. As explained in \cite{Russo:2017ngf,Chen-Lin:2015dfa}, the existence of this parameter makes these Wilson loops good probes of the critical behaviour of the theory.

		The paper is organized as follows. In the next Section, we review the procedure to characterize the phase transitions by studying the large $N$ equations associated to the Chern--Simons-matter matrix model, also introduced in this Section, showing also how the decompactification limit simplifies the equations. This follows the work by Barranco and Russo \cite{Barranco:2014}, with the slight generalization of assuming a non-symmetric interval around the origin for the one-cut solution of the equations. This setting will be used in Section \ref{sect:FItermdistrib}, where we study the phase structure with the presence of a FI parameter, and for a large number of different mass configurations, which leads to a large number of possibilities, which are expounded in detail. In particular for example, we explain how five different phases, instead of the three in \cite{Barranco:2014}, appear due to a ``splitting'' of the masses, caused by the FI parameter. 
		
		In Section \ref{sect:WLmassL} we study Wilson loops in antisymmetric representations for the case without FI parameter. Large $N$ expressions, giving a perimeter law as in \cite{Anderson:2015,Russo:2017ngf}, are obtained for the three phases and we study the finite discontinuity of their second derivative. Various limiting behaviours of the Wilson loops are discussed as well, in Sections \ref{sec:subsecspeciallimit1} and \ref{sec:subsecspeciallimit2}. In Section \ref{sect:oRcorrectionsWL}, $1/R$ corrections are computed for the Wilson loop averages in the different phases. The phase structure is unaffected and the Wilson loop is modified by a shift in the exponential, which is independent of the radius or the 't Hooft parameter. Besides, those corrections do not follow the perimeter law as they introduce a scaling with $k$. In the last Section, we study the Wilson loops in the presence of the FI parameter and with two mass scales, the setting that we had shown, in Section \ref{sect:FItermdistrib}, was characterized by five phase transitions. We compute the five corresponding behaviours for the Wilson loops and find the novel aspect that two transitions are now of first order. The same occurs for one mass scale with one transition of first order, and we explain how this is a consequence of the asymmetry of the density of states in critical points, due to the FI deformation. We finally conclude with some avenues for further research.

	\section{Phase transition with only mass deformation}
	\label{sect:BarrancoRusso}

	\subsection{Chern--Simons-matter theory in the decompactification limit}
		We start our analysis by reviewing the procedure of \cite{Barranco:2014} to obtain the eigenvalue density in the decompactification limit. The model is supersymmetric $U(N)$ Chern--Simons theory on $\cs^3$ with the insertion of two sets of $N_f$ hypermultiplets, with real masses $-m$ and $m$ respectively. In this part of the paper we will be considering the on-shell condition, where the sum of all the masses of the hypermultiplets is zero. However, in later sections we will also show that leaving the masses unconstrained is qualitatively analogous to insert FI deformation. The partition function of the theory localizes to the matrix model \cite{Kapustin:2009kz,Barranco:2014}
		\begin{equation}
		\label{eq:partfunctBarrancoRusso}
			\mathcal{Z} = \int \prod_{i=1} ^N \frac{ \dd x_i }{ 2 \pi} \frac{ \prod_{1\le i < j \le N } \left( 2 \sinh \left( \frac{ x_i - x_j }{2} R \right) \right)^2 }{ \prod_{1 \le i \le N } \left( 2 \cosh \left( \frac{x_i + m}{2} R\right) \right)^{N_f}  \left( 2 \cosh \left( \frac{x_i - m}{2} R \right) \right)^{N_f} }  e^{ - \frac{1}{2 g_s} \sum_{i=1} ^N \left( x_i R \right) ^2 } ,
		\end{equation}
		with $g_s$ the Chern--Simons matrix model coupling, $R$ the radius of $\cs^3$ and the eigenvalues $x_i$ have mass dimension. A comment on notation: the scalar field has mass dimensions, therefore, in the matrix model expression both the integration variable and the mass scale with the radius of the three-sphere. In contrast to \cite{Barranco:2014} and what is customary, we opted to write down the radius $R$ explicitly, since the decompactification limit is central to the whole discussion and ubiquitous in the paper. Indeed, we follow \cite{Barranco:2014} and obtain the eigenvalue density in the decompactification limit $R \to \infty$ with $R$ and $N$ scaling together. The relevant parameters of the theory are
		\begin{equation}
		\label{eq:defparameters}
			t:= g_s N \xrightarrow{N \to \infty} \infty , \quad \lambda := t / m R \equiv \text{ fixed} .
		\end{equation}
		We also assume that $N_f$ scales as $N$, that is, the Veneziano parameter
		\begin{equation}
		\label{eq:defchi}
			\chi := N_f / N 
		\end{equation}
		is kept fixed in the decompactification limit\footnote{In the literature, $N_f /N$ is sometimes referred to with the Greek letter $\zeta$. However, we follow \cite{Anderson:2015} and adopt $\zeta$ for the Fayet--Ilioupoulos parameter, using the Greek letter $\chi$ for the Veneziano parameter.}. Furthermore, we will focus on $0 < \chi < 1$.\par
		We can put the denominator into the exponential in \eqref{eq:partfunctBarrancoRusso} to obtain
		\begin{equation*}
			\mathcal{Z} = \int \prod_{i=1} ^N \frac{ \dd x_i }{ 2 \pi} \prod_{1\le i < j \le N } \left( 2 \sinh \left( \frac{ x_i - x_j }{2} R \right) \right)^2 ~ e^{ - \frac{1}{2 g_s} V (x) } 
		\end{equation*}
		with
		\begin{equation*}
			V (x) := \sum_{i=1 } ^N \left\{ \left( x_i R \right) ^2 + g_s N_f \left[ \log 2 \cosh \left( \frac{ x_i + m }{2} R \right) +  \log 2 \cosh \left( \frac{ x_i - m }{2} R \right) \right] \right\} .
		\end{equation*}
		We so identify the repulsive Vandermonde force and the potential, which is the sum of a harmonic well plus terms which tend to accumulate eigenvalues toward $\pm m$.\par
		For $N \to \infty$ the partition function is given by the saddle points contribution. From the latter form, it is easy to see that, taking derivatives and inserting the eigenvalue density 
		\begin{equation}
		\label{eq:defrhodensity}
			\rho (x) = \frac{1}{N} \sum_{i=1} ^N \delta \left( x - x_i \right) ,
		\end{equation}
		the saddle point equation is written in the integral form \cite{Barranco:2014}
		\begin{equation}
		\label{eq:saddlepointBarrancoRusso}
			\dashint \dd u \rho (u) \coth \left( \frac{ x - u }{2} R \right) - \frac{x}{ m \lambda} - \frac{\chi}{2} \left[ \tanh \left( \frac{ x + m }{2} R \right) + \tanh \left( \frac{ x - m }{2} R \right) \right] = 0 , 
		\end{equation}
		where we used the relations \eqref{eq:defparameters} and \eqref{eq:defchi}, and in particular $g_s N = t = m R \lambda$. The symbol $\dashint$ means the principal value of the integral. In the rest of this section we will find the eigenvalue density and the free energy in the decompactification limit.
		
		\subsubsection{Eigenvalue density}
			We take the decompactification limit of the integral equation \eqref{eq:saddlepointBarrancoRusso}, scaled as explained above. We assume a one-cut solution for $\rho$ \cite{Barranco:2014}, supported in some closed interval $\left[ - A, B \right]$. In the present case with opposite masses and equal number of hypermultiplets associated to each mass, the support will be symmetric, $B=A$. We nevertheless decide to adopt the general formalism since the very beginning.\par
			For $R \to \infty$ the hyperbolic functions in \eqref{eq:saddlepointBarrancoRusso} go to $\pm 1$, depending on the sign of the argument, therefore the saddle point equation significantly simplifies into
			\begin{equation}
			\label{eq:largeRsaddlepointBR}
				\int_{-A} ^B \dd u \rho (u) \mathrm{sign} \left( x - u \right) = \frac{x}{m \lambda} + \frac{\chi}{2} \left[ \mathrm{sign} \left( x + m \right) +  \mathrm{sign} \left( x - m \right) \right] .
			\end{equation}
			The procedure is as follow: depending on the values of $A,B$ the sign functions are to be taken into account or not. Taking the derivative of both sides of \eqref{eq:largeRsaddlepointBR} we are able to find the expression for $\rho$. Then, we impose the normalization condition
			\begin{equation*}
				\int_{-A} ^B \dd x \rho (x) = 1
			\end{equation*}
			which, together with the requirement that $\rho$ must satisfy the full integral expression \eqref{eq:largeRsaddlepointBR}, fixes the values of $A$ and $B$. When one of the boundaries lies on a critical point, its value is known and the pair of additional equations can be used to determine the other boundary and the coefficient of the resonance: this aspect will be clarified below in the case-by-case study.\par
\begin{figure}[hbt]
			\centering
				\includegraphics[width=0.5\textwidth]{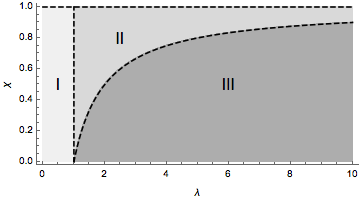}
				\caption{Phase structure for two equal sets of hypermultiplets with only mass deformation.}
				\label{fig:phasediagBaRu}
			\end{figure}

			We now apply the procedure to determine the eigenvalue density and the phase structure, as in \cite{Barranco:2014}.
			\begin{enumerate}[(I)]
				\item The first phase will appear when both $A,B < m$. In this case the sign functions on the r.h.s. of \eqref{eq:largeRsaddlepointBR} cancel out\footnote{Notice that this cancellation happens because both sets yield the same number $N_f$ of hypermultiplets.} and we are left with
					\begin{equation*}
						\int_{-A} ^{x} \dd u \rho (u) - \int_{x} ^B \dd u \rho (u) = \frac{x}{m \lambda} ,
					\end{equation*}
					whose derivative fixes the eigenvalue density to
					\begin{equation*}
						\rho (x) = \frac{1}{2 m \lambda} .
					\end{equation*}
					We now use the normalization condition and impose the full integral equation, using the latter expression for $\rho (x)$, which give
					\begin{equation*}
						 A=B= m \lambda .
					\end{equation*}
					So we obtained a uniform distribution supported in $\left[ - m \lambda , m \lambda \right]$. We underline that the symmetry of the domain is a consequence of the exact cancellation of the sign functions in the r.h.s. of the saddle point equation. The consistency of this phase implies $A,B = m \lambda < m $, that is, such phase appears when $\lambda < 1$.
				\item The next phase appears when the boundaries reach the critical value $A=m $. In this case the sign functions still cancel out in the interior of the domain, but not at the boundary, giving:
					\begin{equation*}
						\rho (x) = \frac{1}{2 m \lambda} + \cII ^{-} \delta \left( x + m \right) + \cII ^{+} \delta \left( x - m \right) ,
					\end{equation*}
					with $\cII ^{\pm}$ coefficients to be fixed. Normalization and integral equation impose
					\begin{equation*}
						 \cII^{-} = \cII^{+} = \frac{ \lambda - 1 }{2 \lambda} ,
					\end{equation*}
					where we have used $A=B=m$.
				\item Eventually we have a phase when the boundaries are greater than the critical value $x=m$. The eigenvalue density is now
					\begin{equation*}
						\rho (x) = \frac{1}{2 m \lambda} + \cIII ^{-} \delta \left( x + m \right) + \cIII ^{+} \delta \left( x - m \right) .
					\end{equation*}
					Moreover, in this phase we have four additional equations: the normalization condition and the integral equation \eqref{eq:largeRsaddlepointBR} for three choices $x< -m$, $-m < x < m$ or $x>m$. So we can determine the full set of four parameters $A,B, \cIII ^{\pm}$ solving the system:
					\begin{equation*}
						\begin{cases} A=B =  m \lambda \left( 1 - \chi \right) , \\ \cIII ^{-} = \cIII^{+} = \frac{\chi}{2} . \end{cases}
					\end{equation*}
					By consistency, this phase holds for $\lambda > \frac{1}{1 - \chi}$.
			\end{enumerate}
			Due to the symmetry of the model, other phases such as $A>m, B<m$ are inconsistent for positive values of $\lambda$ and $\chi$. So we obtained a three-phase structure \cite{Barranco:2014}, sketched in Figure \ref{fig:phasediagBaRu}, with a triple point at $\lambda=1, \chi =0$.\par
			We remark that the partition function and, consequently, the eigenvalue density $\rho$, are invariant under reflection $x_i \mapsto - x_i$.
			
			Before passing to the study of the free energy, we rewrite the eigenvalue density in a compact form
			\begin{equation}
				\rho (x) = \frac{1}{2 m \lambda} + \cp \left[ \delta \left( x + m \right) + \delta \left( x - m \right) \right] , \qquad x \in \left[ - A , A \right]
			\end{equation}
			with $\mathrm{P} \in \left\{ \mathrm{I}, \mathrm{II}, \mathrm{III} \right\}$ and coefficients
			\begin{equation*}
				\cI = 0, \quad \cII = \frac{ \lambda - 1 }{2 \lambda} , \quad \cIII = \frac{\chi}{2} ,
			\end{equation*}
			and, besides, with the expression for $A$ depending on the phase:
			\begin{equation*}
				A \vert_{\mathrm{I}} = m \lambda , \quad A \vert_{\mathrm{II}} = m , \quad A \vert_{\mathrm{III}} = m \lambda \left( 1 - \chi \right) .
			\end{equation*}

		\subsubsection{Free energy}
			We conclude the review of the results of \cite{Barranco:2014} writing down the free energy in each phase. We will see that phase transitions are of third order.\par
			The free energy is defined as
			\begin{equation*}
				\fre = - \frac{1}{R N^2} \log \mathcal{Z} ,
			\end{equation*}
			whose derivative in the large $N$ and large $R$ limit reads:
			\begin{equation*}
				\frac{ \partial \fre }{\partial \lambda} = - \frac{1}{2 m \lambda^2 } \int_{-A} ^B \dd x \rho (x)  x ^2 .
			\end{equation*}
			According to the phase description obtained above, we get
			\begin{equation}
				\frac{ \partial \fre }{\partial \lambda} =  \begin{cases} - \frac{m}{6} , \\  - \frac{m}{6} \left[ 3 \lambda^{-2} - 2 \lambda^{-3}  \right], \\ - \frac{m}{6} \left[ \left( 1 - \chi \right)^3 + 3 \chi \lambda^{-2} \right] , \end{cases} \ \begin{matrix} \ & \lambda  < & 1 , \\ 1 < & \lambda < & \frac{1}{1-\chi} , \\ \ & \lambda  > & \frac{1}{1-\chi} . \end{matrix}
			\end{equation}
			The expression is continuous at both critical lines $\lambda = 1$ and $\lambda = \frac{1}{1 - \chi}$. Taking higher derivatives, it turns out that the first discontinuity appears in the third derivative, that is, the phase transition is of third order. As a concluding remark, we see that for $\chi =0$ phase II is removed, and derivatives of $\fre$ are continuous at the triple point $\lambda=1$ at any order, as it should since in this limit we recover pure Chern--Simons theory on $\cs^3$.\par
			The explicit expression for the free energy is (up to an irrelevant numerical constant):
			\begin{equation}
			\begin{aligned}
				& \fre \vert_{\mathrm{I}}  = - \frac{m}{6} \left[ \lambda - 6 \chi \right] , \\
				& \fre \vert_{\mathrm{II}}  = - \frac{m}{6} \left[ \lambda^{-2} - 3 \lambda^{-1}  + 3 - 6 \chi  \right] , \\
				& \fre \vert_{\mathrm{III}}  = - \frac{m}{6} \left[ \left(1 - \chi \right)^3 \lambda - 3 \chi \lambda^{-1} - 3 \chi \left( 2 - \chi \right) - 6 \chi \right] ,
			\end{aligned}
			\end{equation}
			where we integrated using the fact that, at the saddle points, $\mathcal{Z} (\lambda =0) \approx e^{- N_f N R m }$.

	\section{Phase diagram after insertion of FI term}
	\label{sect:FItermdistrib}
	    
	    After reviewing known methods and results, we proceed now to extend that to a more general setting. Following the procedure of \cite{Barranco:2014,Anderson:2014hxa,Anderson:2015} we obtain the eigenvalue density and the corresponding phase transition in the decompactification limit, when we consider the inclusion of a Fayet--Iliopoulos (FI) term.\par
		We consider two sets of hypermultiplets, where, as before, each set corresponds to a mass scale, but now we let those sets be composed by a number $N_f ^{\pm} \equiv \nu_{\pm} N_f$ of hypermultiplets. Nevertheless, we will first focus on $\nu_{\pm }\in \left\{ 0, 1 \right\}$, that is either only one set or two equal sets, and leave the discussion of the most general case for a later subsection. We assume the masses to be $- m_1$ and $m_2$, and the on-shell condition would imply
		\begin{equation*}
			- \nu_{-}  m_1 + \nu_{+}  m_2 = 0 .
		\end{equation*}
		The partition function is deformed by the insertion of mass and FI term, and localizes into
		\begin{equation*}
			\mathcal{Z} = \int \prod_{i=1} ^N \frac{ \dd x_i}{2 \pi} \frac{ \prod_{1\le i < j \le N} \left( 2 \sinh \left( \frac{ x_i - x_j}{2} R \right) \right)^2 \ e^{- \frac{1}{2 g_s} \sum_{i=1} ^N \left( \left( x_i R \right) ^2 - 2 \zeta R^2 x_i \right) } }{ \prod_{1 \le i \le N } \left( 2 \cosh \left( \frac{ x_i + m_1 }{2} R \right) \right)^{\nu_{-} N_f }  \left( 2 \cosh \left( \frac{ x_i - m_2 }{2} R \right) \right)^{\nu_{+} N_f} }  ,
		\end{equation*}
		where we have denoted by $\zeta$ the --analytically continued-- FI coupling, with mass dimension, as in \cite{Anderson:2014hxa,Anderson:2015}. We see that turning on the FI coupling accounts for a linear term in the exponential. Thus, after a change of variables $x_i \mapsto x_i - \zeta$, this shift is reabsorbed into the mass terms:
		\begin{equation}
		\label{eq:partfunctFIfull}
			\mathcal{Z} = e^{- \frac{ N R^2 \zeta^2  }{2 g_s} } \int \prod_{i=1} ^N \frac{ \dd x_i}{2 \pi} \frac{ \prod_{1\le i < j \le N} \left( 2 \sinh \left( \frac{ x_i - x_j}{2} R \right) \right)^2 \ e^{- \frac{1}{2 g_s} \sum_{i=1} ^N \left( x_i R \right) ^2 } }{ \prod_{1 \le i \le N } \left(  2 \cosh \left( \frac{ x_i + (m_1 - \zeta ) }{2} R \right) \right)^{\nu_{-} N_f} \left( 2 \cosh \left( \frac{ x_i - (m_2 + \zeta ) }{2} R \right) \right)^{\nu_{+} N_f} } .
		\end{equation}
		The FI term therefore produces an effective modification of the mass scales, which now become
		\begin{equation*}
			m_{-} := m_1 - \zeta , \quad m_{+} := m_2 + \zeta .
		\end{equation*}
		We notice that, as a consequence of introducing the FI term, the partition function is no longer invariant under reflection $x_i \mapsto - x_i$, unless one also exchanges the parameters
		\begin{equation*}
			m_{-} \leftrightarrow m_{+} , \quad \nu_{-} \leftrightarrow \nu_{+} .
		\end{equation*}

		\subsection{Decompactification limit with FI term}
		
		We show how the saddle-point analysis works after the insertion of a FI term and in the decompactification limit. We first give the eigenvalue density and then we characterize the free energy.
		
		\subsubsection{Eigenvalue density with FI term}
		
			We now pass to the study of the eigenvalue density in the decompactification limit $R \to \infty$ with $\lambda= t/m R$ fixed.\par
			We rewrite the denominator of the partition function \eqref{eq:partfunctFIfull} in the exponential:
			\begin{equation*}
%			\label{eq:FIpartfunctpotential}
				\mathcal{Z} = e^{- \frac{ N R^2 \zeta^2 }{2 g_s} } \int \prod_{i=1} ^N \frac{ \dd x_i}{2 \pi} \prod_{1\le i < j \le N} \left( 2 \sinh \left( \frac{ x_i - x_j}{2} R \right) \right)^2 \ e^{- \frac{1}{2 g_s} V (x) } ,
			\end{equation*}
			where
			\begin{equation*}
				V(x) = \sum_{i=1} ^N \left\{ (R x_i) ^2 + 2 g_s N_f \left[ \nu_{-} \log \left( 2 \cosh \left( \frac{ x_i + m_{-} }{2} R \right) \right)  +  \nu_{+} \log \left( 2 \cosh \left( \frac{ x_i - m_{+} }{2} R \right) \right) \right] \right\} .
			\end{equation*}
			As usual, we have a repulsive Vandermonde force, which tends to spread the eigenvalues apart, and a potential composed by a harmonic well plus two attractive terms which tend to cluster eigenvalues around the resonances at $\pm m_{\pm}$. The saddle point contribution is to be determined solving the following system of equations, one for each eigenvalue:
			\begin{equation*}
				\sum_{j \ne i}  \coth \left( \frac{ x_i - x_j}{2} R \right) - \frac{1}{2g_s} \left\{ 2 x_i + g_s N_f  \left[ \nu_{-} \tanh \left( \frac{ x_i + m_{-} }{2} R \right) + \nu_{+} \tanh \left( \frac{ x_i - m_{+} }{2} R \right) \right] \right\} = 0 ,
			\end{equation*}
			for $i=1, \dots, N$. We now introduce the eigenvalue density as in \eqref{eq:defrhodensity}
			\begin{equation*}
				\rho (x) := \frac{1}{N} \sum_{i=1} ^N \delta (x - x_i) , \qquad x \in \left[ - A , B \right] ,
			\end{equation*}
			assuming one-cut solution in analogy with \cite{Anderson:2015}. It is important to notice that the FI term explicitly breaks the reflection symmetry, even when we chose sets with equal number of hypermultiplets $\nu_{-} = \nu_{+}$, therefore one should expect non-symmetric domain $A \ne B$. Plugging the eigenvalue density reduces the system of $N$ saddle point equations into the unique integral equation
			\begin{equation*}
				\dashint_{-A} ^{B} \dd y \rho (y) \coth \left( \frac{ x - y}{2} R \right)  - \frac{x }{m \lambda } - \frac{ \chi}{2} \left[ \nu_{-} \tanh \left( \frac{ x + m_{-} }{2} R \right) + \nu_{+} \tanh \left( \frac{ x - m_{+} }{2} R \right) \right]  = 0 ,
			\end{equation*}
			where $\dashint$ means the principal value of the integral, and we used the definition of the suitable parameters \eqref{eq:defparameters} and \eqref{eq:defchi} to substitute $g_s N = t = m R \lambda$. We have introduced a mass scale $m$ given by the average of the masses
			\begin{equation*}
				m := \sum_{a} \frac{\nu_{a}}{\sum_{a} \nu_{a}} m_a ,
			\end{equation*}
			in order to properly and uniquely define the parameter $\lambda$. Clearly, setting $\nu_{-} = \nu_{+} = 1$, which corresponds to have equal number of hypermultiplets associated to both mass scales, and imposing the on-shell constraint $m_1 = m_2 \equiv m$, we recover equation \eqref{eq:saddlepointBarrancoRusso} in the limit $\zeta \to 0$. Thus, nonvanishing $\zeta$ deforms the theory splitting the mass values into two effective masses $m_{\pm}$.\par
			In the decompactification limit the hyperbolic functions reduce to sign functions, and the saddle point equation is then
			\begin{equation}
			\label{eq:largeRsaddleeq}
				\int_{- A} ^{B} \dd y \rho (y) \mathrm{sign} \left( x - y \right) = \frac{x}{m \lambda} + \frac{ \chi }{2} \left[ \nu_{-} \mathrm{sign} \left( x + m_{-} \right) + \nu_{+} \mathrm{sign} \left( x - m_{+} \right) \right]  .
			\end{equation}
			The procedure to solve it is as follow: we identify different phases according to the values of the boundaries $A,B$ of the domain, then in each phase we take the derivative of the saddle point equation \eqref{eq:largeRsaddleeq} to obtain the expression for $\rho (x)$, and finally we take advantage of the normalization condition and of the full integral expression \eqref{eq:largeRsaddleeq} to fix the values of the parameters. For later convenience we report here the general form of the solution:
			\begin{equation}
			\label{eq:rhogenP}
				\rho (x) = \frac{1}{2 m \lambda} + \cp ^{-} \delta \left( x + m_{-} \right) + \cp ^{+} \delta \left( x - m_{+} \right) , \qquad x \in \left[ - A, B \right] ,
			\end{equation}
			where $\mathrm{P}$ labels the phase. One always finds $\cI ^{\pm} =0$ in the first phase and, in the last phase, $C_{\mathrm{last}} ^{\pm} = \nu_{\pm} \frac{ \chi}{2} $.\par
			To proceed further, we split our analysis in two sub-cases \cite{Anderson:2015}, namely $0<\zeta<m_1$ and $\zeta \ge m_1$. Without loss of generality we avoid the discussion $\zeta < 0$ because, as mentioned before, the reflection symmetry allows us to recover this case, up to relabelling $\nu_{\pm} \mapsto \nu_{\mp}$ and $m_{\pm} \mapsto m_{\mp}$. Furthermore, we will first focus on the most interesting cases $\nu_{\pm} \in \left\{ 0 , 1 \right\}$, while a more general analysis will be given later.

		\subsubsection{Free energy with FI term}
		
			Once the explicit expression for the eigenvalue distribution $\rho$ is known in each phase, one is able to calculate the free energy. In the large $N$ limit the partition function is given exactly by its saddle point contribution, which is encoded in $\rho$. Consider
			\begin{equation*}
				\fee = - \frac{1}{R N^2} \log \mathcal{Z} - \frac{ \zeta^2 R }{2 g_s N} ,
			\end{equation*}
			the free energy shifted not to keep track of the constant term. One obtains the formula for its derivative at large $N$:
			\begin{equation}
				\frac{ \partial \fee }{\partial \lambda} = - \frac{1}{2 m \lambda^2} \int_{-A} ^B \dd x \rho (x)  x ^2 .
			\end{equation}
			We can evaluate $\frac{ \partial \fee }{\partial \lambda}$ in each phase inserting the corresponding expression for $\rho$. From \eqref{eq:rhogenP} we obtain the general expression
			\begin{equation}
			\label{eq:generalfreeenergy}
				\frac{ \partial \fee }{\partial \lambda} = - \frac{1}{2 m \lambda^2} \left[ \frac{A ^3 + B ^3 }{6 m \lambda } + \cp ^{-} m_{-} ^2 + \cp ^{+} m_{+} ^2 \right] .
			\end{equation}

		\subsection{$ \zeta < m_1$ with two mass scales}
		\label{sect:rhotwofamFI}
		
			\subsubsection{Eigenvalue density for $ \zeta < m_1$ and $\nu_{-} = \nu_{+} = 1$}
				We first investigate the case $0<\zeta < m_1$, which guarantees $m_{-} > 0$. Under those circumstances, the model presents a resonance at negative value $- m_{-}$ and one at positive value $m_{+}$. Furthermore we set $\nu_{-} = \nu_{+} = 1$. This choice allows for a simpler analysis and can also be compared to the results of \cite{Barranco:2014} reviewed in the previous Section \ref{sect:BarrancoRusso}, in the limit $\zeta \to 0$. Moreover, we assume for simplicity that the ordering $m_{-} < m_{+}$ holds, which is indeed satisfied if we impose the on-shell constraint $m_1=m_2 \equiv m$. The integral saddle point equation \eqref{eq:largeRsaddleeq} with those hypotheses becomes
				\begin{equation}
				\label{eq:integraleq2fam}
					\int_{-A} ^{x} \dd y \rho (y)  - \int_{x} ^{B} \dd y \rho (y) =  \frac{x}{m \lambda} + \frac{\chi}{2} \left[ \mathrm{sign} \left( x + m_{-} \right) + \mathrm{sign} \left( x - m_{+} \right) \right] .
				\end{equation}
				\begin{enumerate}[(I)]
					\item The first phase appears when both resonances are placed out of the domain: $A< m_{-} $ and $B< m_{+}$. Taking the derivative of equation \eqref{eq:integraleq2fam} we find the uniform distribution
						\begin{equation}
							\rho (x) = \frac{1}{2 m \lambda} .
						\end{equation}
						We now plug this expression into the normalization condition $\int_{-A} ^B \dd x \rho (x) = 1$ and into the integral equation \eqref{eq:integraleq2fam} to determine the values of $A$ and $B$. This gives:
						\begin{equation*}
							A=B=m \lambda.
						\end{equation*}
						In particular, for $\nu_{-} = \nu_{+}$ the sign functions on the r.h.s. of the saddle point equation have same coefficient and cancel out exactly, leading to symmetric domain $A=B$. Since $m_{-} < m_{+}$, the consistency of this phase requires $A= m \lambda < m_{-}$, that is, this first phase corresponds to
						\begin{equation*}
							\lambda < \frac{m_{-}}{m} .
						\end{equation*}
						We notice that the splitting of the mass value due to the presence of the FI coupling $\zeta$ moves the critical value away from $\lambda=1$.
					\item The next phase appears when, increasing $\lambda$, one of the boundaries reaches a resonance. As we are assuming the on-shell ordering $m_{-} < m_{+}$ to hold, the second phase appears when $A = m_{-} + \oR$ and $B< m_{+}$. Only one of the sign functions contributes, and only at the boundary. Hence
						\begin{equation}
							\rho (x) = \frac{1}{2 m \lambda} + \cII ^{-} \delta \left( x + m_{-} \right) ,
						\end{equation}
						with $\cII^{-}$ and $B$ to be determined using normalization and the full integral expression \eqref{eq:integraleq2fam}. Those two equations provide:
						\begin{equation*}
							 \cII^{-} = \frac{ \lambda - m_{-}/m }{2 \lambda} , \quad  B= m \lambda ,
						\end{equation*}
						where we have used $A=m_{-}$. The appearance of a resonance at one boundary does not modify the behaviour of the other boundary.
					\item At this point two possibilities disclose: either $A$ increases and overcomes $m_{-}$ while $B$ is still below the critical value $m_{+}$, or instead $A$ is kept at the resonance value and $B$ reaches $m_{+}$.
						\begin{enumerate}[(a)]
							\item $A> m_{-}$ and $B< m_{+}$. The derivative of \eqref{eq:integraleq2fam} gives
								\begin{equation}
									\rho (x) = \frac{1}{2 m \lambda} + \cIII^{-} \delta \left( x + m_{-} \right) ,
								\end{equation}
								and we determine $A,B,\cIII^{-}$ using the normalization condition together with the integral expression, for two different choices $x < - m_{-}$ or $x> m_{-}$, obtaining
								\begin{equation*}
									 \cIII^{-} = \frac{ \chi }{2 } , \quad  A= m \lambda \left( 1 - \chi \right) , \quad B= m \lambda .
								\end{equation*}
								This phase holds as long as $A= m \lambda \left( 1 - \chi \right) > m_{-}$ and $B= m \lambda < m_{+}$, thus
								\begin{equation*}
									\frac{m_{-}}{m \left( 1 - \chi \right)} < \lambda < \frac{m_{+} }{m} .
								\end{equation*}
								This is consistent only if $\frac{m_{-}}{1 - \chi } < m_{+}$, that is, whenever $\chi$ is below a critical value
								\begin{equation*}
									\chi < 1 - \frac{m_{-}}{m_{+}} ,
								\end{equation*}
								which in the on-shell case $m_1=m_2=m$ is given by $\frac{2 \zeta}{m + \zeta}$.
							\item $A= m_{-} + \oR$ and $B = m_{+} + \oR$. In this case both resonance play a role, with the eigenvalue density given by
								\begin{equation}
									\rho (x) = \frac{1}{2 m \lambda} + \cIII ^{-} \delta \left( x + m_{-} \right) +  \cIII ^{+} \delta \left( x - m_{+} \right) ,
								\end{equation}
								\begin{equation*}
									 \cIII^{-} = \frac{ \lambda - m_{-}/m }{2 \lambda} , \quad  \cIII^{+} = \frac{ \lambda - m_{+}/m }{2 \lambda} .
								\end{equation*}
								This phase appears for
								\begin{equation*}
									 \frac{m_{+}}{m} < \lambda < \frac{m_{-}}{m \left( 1 - \chi \right)} .
								\end{equation*}
								Of course, this is possible only when $\chi > 1 - \frac{m_{-}}{m_{+}}$, thus this case is complementary to the previous one.
						\end{enumerate}
					\item The following phase arises for $A>m_{-}$ and $B = m_{+} + \oR$, independently of which of the two cases is realized for phase III. In this phase, the negative resonance is in the interior of the domain while the positive one lies on the boundary. The eigenvalue density reads:
						\begin{equation}
							\rho (x) = \frac{1}{2 m \lambda} + \cIV ^{-} \delta \left( x + m_{-} \right) +  \cIV ^{+} \delta \left( x - m_{+} \right) .
						\end{equation}
						We now have three available equations: normalization and integral expression for $x< -m_{-}$ and for $x>- m_{-}$. We thus determine
						\begin{equation*}
							 \cIV^{-} = \frac{ \chi }{2 } , \quad \cIV ^{+} = \frac{ \lambda - m_{+}/m }{2 \lambda} , \quad A= m \lambda \left( 1 - \chi \right) . 
						\end{equation*}
						This phase appears when
						\begin{equation*}
							\lambda > \max \left\{ \frac{m_{-}}{m \left( 1 - \chi \right)} , \frac{m_{+}}{m} \right\} .
						\end{equation*}
					\item We have a last phase, when both boundaries are such that the resonances lie in the interior of the domain, which means $A> m_{-}$ and $B> m_{+}$. In this case, the eigenvalue density is:
						\begin{equation}
							\rho (x)= \frac{1}{2 m \lambda} + \cV ^{-} \delta \left( x + m_{-} \right) +  \cV ^{+} \delta \left( x - m_{+} \right)
						\end{equation}
						and now we have four equations to determine all the parameters $A,B, \cV ^{\pm}$, which correspond to normalization and integral equation for $x < - m_{-} $, $- m_{-} < x < m_{+}$ or $x> m_{+}$. Those lead to:
						\begin{equation*}
							\begin{cases} \cV^{-} = \cV^{+}  = \frac{\chi }{ 2 } ,  \\ A = B = m \lambda \left( 1 - \chi \right) . \end{cases}
						\end{equation*}
						The consistency of this phase requires that $B=m \lambda \left( 1 - \chi \right)$ overcomes the critical point $m_{+}$, hence phase V arises at
						\begin{equation*}
							\lambda > \frac{ m_{+} }{m \left( 1 - \chi \right) } .
						\end{equation*}
				\end{enumerate}\par
\begin{figure}[hbt]
					\centering
					\includegraphics[width=0.5\textwidth]{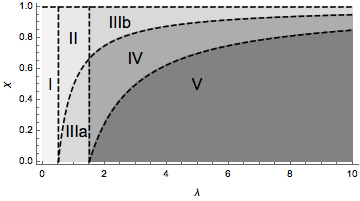}
					\caption{Phase structure with two sets with equal number of hypermultiplets, for $\zeta = m/2$. We see five phases, with third phase different according to $\chi $ above or below the value $ 1 - m_{-}/m_{+}$.}
				\label{fig:phasestruct2fam}
				\end{figure}

				\medskip
				Therefore we have five phases: phase transitions of Section \ref{sect:BarrancoRusso} are split into two. This happens even in the on-shell case $m_1=m_2=m$: the FI coupling splits the mass into $m_{\pm}$, which implies a splitting of the critical curves. The phase structure in the on-shell case is presented in Figure \ref{fig:phasestruct2fam}. In particular, we see how, for a certain value of $\chi$, two critical lines cross and there is a switch in phase III. We also notice that the triple point $\lambda = 1$ splits into $\lambda = \frac{ m_{\pm} }{m}$. Taking the limit $\zeta \to 0$ the curves converge to each other and eventually the result \cite{Barranco:2014} of Section \ref{sect:BarrancoRusso} is recovered, with its three phases.
				
				\par
				\medskip
				We notice that, as long as the ordering $m_{-} < m_{+}$ holds, there is no difference between on-shell and unconstrained case. In fact, the masses are already split by the FI term, and adding further difference does not change the phase structure. In the off-shell case, however, it could happen\footnote{This corresponds to $m_1 >m_2$ with $0<\zeta< \frac{m_1 - m_2}{2}$.} that $m_{-} > m_{+}$: if this is the case, the five-phase structure persists, but the role of $A$ and $B$ is exchanged. So, for instance, the first transition would happen because $B$ reaches $m_{+}$ while $A$ is below its critical value $m_{-}$. We conclude that the unconstrained case is qualitatively analogous to the physical one.

			\subsubsection{Free energy for $ \zeta < m_1$ and $\nu_{-} = \nu_{+} = 1$}
				We now study the free energy and determine the order of the phase transition under those circumstances. We will only have to insert in the expression \eqref{eq:generalfreeenergy} the values of $A,B$ and of the coefficients in each phase, as obtained above. We divide the study into two sub-cases, depending on whether $\chi< 1 - \frac{m_{-}}{m_{+}}$ or not.\par
				For the case $\chi < 1 - \frac{m_{-} }{ m_{+}}$ we get:\begingroup
\renewcommand*{\arraystretch}{1.2}
				\begin{equation}
				\label{eq:firstorderFEa1}
					\frac{ \partial \fee }{ \partial \lambda } = \begin{cases} - \frac{m}{6} , \\  - \frac{m}{12 } \left[1 + 3 \lambda^{-2}  \left( \frac{ m_{-} }{m} \right)^2 - 2  \lambda^{-3} \left( \frac{ m_{-} }{m} \right)^3 \right] ,  \\ - \frac{m}{12 } \left[ 1 + \left(1- \chi \right)^3   + 3 \chi \lambda^{-2} \left( \frac{ m_{-} }{m} \right)^2 \right],  \\  - \frac{m}{12 } \left[ \left(1- \chi \right)^3  + 3 \lambda^{-2}  \left(  \frac{ \chi m_{-}^2 + m_{+}^2  }{m^2} \right) - 2 \lambda^{-3} \left( \frac{ m_{+} }{m} \right)^3 \right] ,  \\   - \frac{m}{12 } \left[  2 \left(1- \chi \right)^3  + 3 \chi \lambda ^{-2} \left(  \frac{m_{-} ^2 + m_{+}^2 }{m^2}  \right) \right],  \end{cases} \ \begin{matrix} \ & \quad \lambda < & \frac{ m_{-} }{m} , \\ \frac{ m_{-} }{m} & < \lambda < & \frac{m_{-}}{m \left( 1 -  \chi \right) } , \\  \frac{m_{-}}{m \left( 1 -  \chi \right)} & <  \lambda < & \frac{ m_{+}}{m} , \\ \frac{m_{+}}{m} & <  \lambda < & \frac{m_{+}}{m \left( 1 - \chi \right)} , \\ \ & \quad \lambda > & \frac{ m_{+} }{ m \left( 1 - \chi \right)} .  \end{matrix}
				\end{equation}\endgroup
				This is a continuous function at all critical points. When $\chi > 1 - \frac{ m_{-}}{m_{+}}$, the expressions in all phases but the third remain the same, with upper range boundary of phase II and lower range boundary of phase IV exchanged. In formulae:\begingroup
\renewcommand*{\arraystretch}{1.3}
				\begin{equation}
				\label{eq:firstorderFEa2}
					\frac{ \partial \fee }{ \partial \lambda } = \begin{cases} - \frac{m}{6} ,  \\  - \frac{m}{12 } \left[1 + 3 \lambda^{-2}  \left( \frac{ m_{-} }{m} \right)^2 - 2  \lambda^{-3} \left( \frac{ m_{-} }{m} \right)^3 \right],  \\ - \frac{m}{12} \left[ 3 \lambda^{-2} \left( \frac{m_{-}^2 + m_{+}^2 }{m^2} \right) - 2 \lambda^{-3} \left( \frac{ m_{-}^3 + m_{+} ^3 }{m^3} \right) \right] ,  \\  - \frac{m}{12 } \left[ \left(1- \chi \right)^3  + 3 \lambda^{-2}  \left(  \frac{ \chi m_{-}^2 m_{+} ^2 }{m^2}  \right) - 2 \lambda^{-3} \left( \frac{ m_{+} }{m} \right)^3 \right] ,  \\   - \frac{m}{12 } \left[ 2  \left(1- \chi \right)^3  + 3 \chi \lambda^{-2}\left( \frac{ m_{-}^2 + m_{+}^2 }{m^2} \right) \right],  \end{cases}  \ \begin{matrix} \ & \quad \lambda < & \frac{ m_{-} }{m} , \\ \frac{ m_{-} }{m} & < \lambda < & \frac{m_{+}}{m} , \\  \frac{ m_{+}}{m}  & <  \lambda < & \frac{m_{-}}{m \left( 1 -  \chi \right)} , \\  \frac{m_{-}}{m \left( 1 -  \chi \right) } & <  \lambda < & \frac{m_{+}}{m \left( 1 - \chi \right)} , \\ \ & \quad \lambda > & \frac{ m_{+} }{ m \left( 1 - \chi \right)} .  \end{matrix}
				\end{equation}\endgroup
				where the evaluation in phase III may appear in a simpler form using the on-shell relations $ m_{-} ^2 + m_{+} ^2 = 2m^2 \left( 1 + \frac{ \zeta^2}{m^2} \right)$ and $m_{-} ^3 + m_{+} ^3 = 2 m^3 \left( 1 + 3 \frac{ \zeta^2 }{m^2}  \right) $. This is again a continuous function. We underline that this is the case to consider in the FI-free limit $\zeta \to 0$, and in that limit we recover the results of Section \ref{sect:BarrancoRusso}.\par
				To obtain the order of the phase transition we have to take further derivatives: this is done explicitly in Appendix \ref{app:freeen2d}. We have that second derivatives are continuous, while third derivatives are not. Hence, all phase transitions are of third order, consistently with \cite{Barranco:2014,Anderson:2015}.\par
				We conclude providing $\fee$ explicitly phase by phase, when $\chi > 1 - \frac{m_{-}}{m_{+}}$:
				{\small\begin{equation}
				\begin{aligned}
					& \fee \vert_{\mathrm{I}}  = - \frac{m}{6} \left[ \lambda - 6 \chi \right] , \\
					& \fee \vert_{\mathrm{II}}  = - \frac{m}{12} \left[ \lambda - 3 \lambda^{-1} \left( \frac{m_{-}}{m} \right)^2 + 2 \lambda^{-2} \left( \frac{m_{-}}{m} \right)^3 + 2 \left( \frac{m_{-}}{m} \right) - 12 \chi \right] , \\
					& \fee \vert_{\mathrm{III}} = - \frac{m}{12} \left[ - 3 \lambda^{-1} \left( \frac{ m_{-}^2 + m_{+}^2}{m^2} \right) + \lambda^{-2} \left( \frac{ m_{-}^3 + m_{+}^3}{m^3}  \right) + \frac{m_{-}}{m} \left( 2 + \frac{ m_{-}^2}{m_{+} ^2} \right) + 3 \frac{m_{+}}{m} - 12 \chi \right] , \\
					& \fee \vert_{\mathrm{IV}} = - \frac{m}{12} \left[ \left( 1 - \chi \right)^3 \lambda - 3 \lambda^{-1} \left(  \frac{ \chi m_{-}^2 + m_{+}^2}{m^2}\right) + \lambda^{-2} \frac{m_{+} ^3 }{m^3} +  \frac{m_{-}}{m} \left( -1 + 3 \chi \left( 2 - \chi \right) + \frac{ m_{-}^2}{m_{+} ^2} \right) + 3 \frac{m_{+}}{m} - 12 \chi \right] , \\
					& \fee \vert_{\mathrm{V}} = - \frac{m}{12} \left[  2 \left( 1 - \chi \right)^3 \lambda  - 3 \chi  \lambda^{-1} \left( \frac{m_{-} ^2 + m_{+} ^2 }{m^2} \right) +  \frac{m_{-}}{m} \left( -1 + 3 \chi \left( 2 - \chi \right) + \frac{ m_{-}^2}{m_{+} ^2} \right) + 3 \chi \left( 2 - \chi \right)\frac{m_{+}}{m}  - 12 \chi \right] .
				\end{aligned}
				\end{equation}}

		\subsubsection{Decompactification limit at finite $N$}
		\label{sec:decompactfiniteN}
			The behaviour of the partition function \eqref{eq:partfunctBarrancoRusso} in the decompactification limit keeping $N$ finite was analyzed in \cite{Russo:2014bda} using the equivalent representation of \eqref{eq:partfunctBarrancoRusso} as a determinant of a matrix whose entries are Mordell integrals. It was still possible to identify the three phases at finite $N$, due to the presence of the additional parameter $R$ which was sent to infinity. We shortly review this here. Let us start from the identification $\mathcal{Z} = \mathcal{C} \det J_{jk}$, with $1 \le j,k \le N$ and $\mathcal{C}$ an irrelevant overall constant, and
			\begin{equation*}
				J_{jk} = \int_{- \infty} ^{+ \infty} \dd x \frac{ e^{- \frac{g_s}{2} ( (xR)^2 -2x R(j+k-1-N)) } }{ [4 \cosh \left( \frac{ g_s x + m}{2} R \right) \cosh \left( \frac{ g_s x - m}{2} R\right) ]^{N_f}} .
			\end{equation*}
			For shortness, we will henceforth denote $l := j+k-1-N$, with $\lvert l \rvert \le N-1$, and $m= g_s p$. Approximating the hyperbolic functions in the denominator as $R \to \infty$, we get:
			\begin{equation*}
				J_{jk} \approx \int_{- \infty } ^{-p} \dd x e^{ - \frac{g_s}{2} ((xR)^2 - 2 xR (l + N_f)) } + \int_{-p} ^{p} \dd x e^{- \frac{g_s}{2} ((xR)^2 - 2 xR l + 2 p N_f)) } + \int_{p} ^{+ \infty} \dd x e^{ - \frac{g_s}{2} ((xR)^2 - 2 xR (l - N_f)) } .
			\end{equation*}
			Whether the first, second or third integral gives the leading contribution depends on which interval contains the saddle point. If $N-1 < p$, the saddle point is $xR = l$ and lies in $(-p,p)$ and the leading contribution comes from the second integral. If\footnote{Recall that we are working in the regime $N_f < N$.} $N-1-N_f < p < N-1$, the leading contribution comes from the boundaries $x R = \pm p$. Finally, when $N-1-N_f > p$, the leading contribution comes from the saddle points at $l \pm N_f$, and depending on the value of $l$ (hence of $j,k$) it lies in the domain of the first or third integral. See \cite{Russo:2014bda} for further details and the computation of $J_{jk}$ in each phase.\par
			The same procedure applies here. We describe the finite $N$, large $R$ regime in the case of two mass scales and FI term, and the procedure can be extended straightforwardly to the other cases. We have:
			\begin{equation*}
				J_{jk} = \int_{- \infty} ^{+ \infty} \dd x \frac{ e^{- \frac{g_s}{2} ( (xR)^2 -2x R l ) } }{ [4 \cosh \left( \frac{ g_s x + m_{-} }{2} R \right) \cosh \left( \frac{ g_s x - m_{+} }{2} R\right) ]^{N_f}} ,			
			\end{equation*}
			and, denoting $m_{\pm} = g_s p_{\pm}$, the decompactification limit splits again the integral into:
			\begin{equation*}
			\begin{aligned}
				J_{jk} & \approx \int_{- \infty } ^{-p_{-}} \dd x e^{ - \frac{g_s}{2} \left( (xR)^2 - 2 xR (l + N_f) + (p_{+} - p_{-}) N_f \right) } + \int_{-p_{-}} ^{p_{+}} \dd x e^{- \frac{g_s}{2} ((xR)^2 - 2 xR l + (p_{+} + p_{-}) N_f)) } \\
				& + \int_{p_{+}} ^{+ \infty} \dd x e^{ - \frac{g_s}{2} ((xR)^2 - 2 xR (l - N_f) - (p_{+}-p_{-}) N_f) } .
			\end{aligned}
			\end{equation*}
			For $N-1<p_{-}$ the saddle point falls into $(- p_{-}, p_{+})$ and the leading contribution comes from the second integral, while for $N-1-N_f > p_{+}$ the leading contribution comes either from the first or third integral, depending on $l$. The difference now is that the middle region can receive contributions either from one of the boundaries, or both. In fact, for $N-1-N_f < p_{-}<N-1<p_{+}$, the contributions come from the boundary at $-p_{-}$ and the second integral. Conversely, for $p_{-} < N-1-N_f < p_{+} < N-1$, the contributions come from the first integral and the boundary $p_{+}$. Finally, we see that, for the middle phase, two scenarios are possible: either $p_{-}<N-1-N_f<N-1<p_{+}$, and the saddle points lie in the domain of the first and second integral, or $N-1-N_f < p_{-} < p_{+} < N-1$ and the saddle points lie at the boundaries $\pm p_{\pm}$. We can still see the five-phase structure, and the critical values are in agreement with those found at large $N$.\par

		\subsection{Other cases}
		\label{sect:zsmall1fam}
		
\begin{figure}[hbt]
						\centering
						\includegraphics[width=0.5\textwidth]{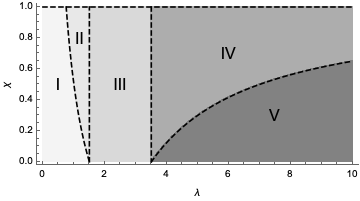}
						\caption{Phase structure with two sets with equal number of hypermultiplets, for $\zeta = 5m/2$. Two mass scales produce five phases, with unique third phase a stripe of width $2$.}
					\label{fig:phasediag2famzetabigger}
				\end{figure}

		Here we present the phase structure and the free energies in other cases. In particular, we focus on the cases of two equal sets of hypermultiplets $\nu_{-} = \nu_{+}$ and $m_1 < \zeta$ (Figure \ref{fig:phasediag2famzetabigger}) and only one set of hypermultiplets with mass $-m_1 < -\zeta$ (Figure \ref{fig:phasediag1famFI}). The eigenvalue densities are reported in Appendix \ref{app:tablerho}. Other cases as, for example, $\nu_{-}=0,\nu_{+}=1$, can be easily retrieved by one of the cases illustrated here. We notice that, for the two-mass case with $0<-m_{-}<m_{+}$, the strict inequalities guarantee the absence of crossings of critical lines, hence the middle phase in uniquely determined, and shaped as a stripe of width $\frac{m_{+} + m_{-}}{m} =2$ (Figure \ref{fig:phasediag2famzetabigger}).
		
				\begin{figure}[hbt]
					\centering
					\includegraphics[width=0.5\textwidth]{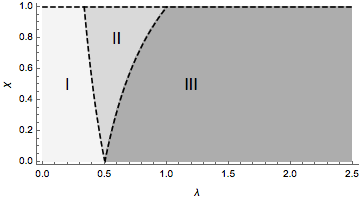}
					\caption{Phase structure with one set of hypermultiplets, for $\zeta = m/2$. Only one mass scale generates three phases.}
				\label{fig:phasediag1famFI}
				\end{figure}\par

				From the general expression \eqref{eq:generalfreeenergy} we obtain the free energy for the case $\nu_{-} =1= \nu_{+}$:
					{\small\begin{equation}
				\begin{aligned}
					& \fee \vert_{\mathrm{I}} = - \frac{m}{6} \left[ \left( 1 + 3 \chi^2 \right) \lambda - 6 \chi \right] , \\
					& \fee \vert_{\mathrm{II}} = - \frac{m}{12} \left[ \left( 1 - \chi \right)^3 \lambda -3 \left(1 + \chi \right) \lambda^{-1}  \left( \frac{ m_{-}^2 }{m^2} \right) + \lambda^{-2} \left( \frac{ - m_{-}^3 }{m^3} \right) + 3 \left( 1 + \chi \right)^2 \left( \frac{ - m_{-} }{m} \right) - 12 \chi \right] , \\
					& \fee \vert_{\mathrm{III}} = - \frac{m}{12} \left[ \left( 1 + \left( 1 - \chi \right)^3 \right) \lambda  - 3 \chi \lambda^{-1} \left( \frac{m_{-}^2 }{m^2} \right) + 3 \chi \left( 2 - \chi \right) \left( \frac{ - m_{-} }{m} \right) - 12 \chi \right] , \\ 
					& \fee \vert_{\mathrm{IV}} = - \frac{m}{12} \left[  \left( 1 - \chi \right)^3 \lambda - 3 \lambda^{-1} \left( \frac{ \chi m_{-} ^2 + m_{+}^2 }{m^2} \right) + \lambda^{-2} \left( \frac{ m_{+} ^3 }{m^3}\right) + 3 \left(\frac{m_{+}}{m}\right) + 3 \chi \left( 2 - \chi \right) \left( \frac{ - m_{-} }{m} \right) - 12 \chi \right] , \\
					& \fee \vert_{\mathrm{V}} = - \frac{m}{6} \left[ \left( 1 - \chi \right)^3 \lambda - \frac{3 \chi}{2} \lambda^{-1}  \left( \frac{ m_{-} ^2 + m_{+}^2 }{m^2} \right) + 3 \chi \left( 2 - \chi \right)  - 6 \chi \right] ,
				\end{aligned}
				\end{equation}}
				while for $\nu_{-}=1, \nu_{+}=0$ and $\zeta>m_1$ it leads to:
				{\small\begin{equation}
				\begin{aligned}
					& \fee \vert_{\mathrm{I}} = - \frac{m}{6} \left[ \lambda \left( 1 + 3 \left( \frac{ \chi}{2}\right)^2 \right) - 3 \chi \right] , \\
					& \fee \vert_{\mathrm{II}} = - \frac{m}{12} \left[ \left( 1 - \frac{\chi}{2} \right)^3 \lambda -3 \left( 1 + \frac{\chi}{2} \right)\lambda^{-1} \frac{ m_{-}^2}{m^2} + \lambda^{-2} \frac{ m_{-}^3}{m^3} + 3 \left( 1 + \frac{\chi}{2} \right)^2 \frac{m_{-}}{m} - 6 \chi \right] ,  \\
					& \fee \vert_{\mathrm{I}} = - \frac{m}{6} \left[ \left( 1 - \frac{\chi}{2} \right)^3 \lambda - \frac{ 3 \chi}{2}  \lambda^{-1} \frac{ m_{-}^2}{m^2}  + \chi \left( 1 + \frac{\chi}{2}\right)\frac{m_{-}}{m} - 3 \chi  \right] , 
				\end{aligned}
				\end{equation}}
				where we used the boundary condition $\widetilde{\mathcal{Z} }(\lambda =0) \approx e^{-N_f N R m/2}$. All phase transitions are third order, see Appendix \ref{app:freeen2d} for calculations.\par
				A finite $N$ analysis with $R \to \infty$ can be done, in exactly the same manner as in Subsection \ref{sec:decompactfiniteN}. For example, for the case of a single mass scale, the integrals $J_{jk}$ split into two and, depending on the value of the mass, the saddle point lies in the domain of the first or second integral or at the boundary.

		\subsection{A generic number $S$ of mass scales}
			One could generalize the previous setting by picking $S$ sets of massive hypermultiplets, each set characterized by mass $\epsilon (a) m_a$ and multiplicity $N_{f,a}$, with $\epsilon (a) = \pm 1$, for $a=1, \dots, S$. The partition function would be
			\begin{equation}
				\mathcal{Z} = \int \prod_{i=1} ^N \frac{ \dd x_i}{2 \pi} \frac{ \prod_{1\le i < j \le N} \left( 2 \sinh \left( \frac{ x_i - x_j}{2} R \right) \right)^2 \   e^{- \frac{1}{2 g_s} \sum_{i=1} ^N \left( R^2 x_i ^2 - 2 \zeta R^2  x_i \right) } }{ \prod_{1 \le i \le N } \prod_{a=1} ^{S} \left( 2 \cosh \left( \frac{ x_i - \epsilon (a) m_a }{2} R \right) \right)^{ N_{f,a} } } ,
			\end{equation}
			The on-shell condition in this generalized framework is:
			\begin{equation}
				\sum_{a=1} ^S N_{f,a} \epsilon (a) m_a = 0 .
			\end{equation}
			To define the 't Hooft parameter and its scaling in the decompactification limit, we introduce the average mass
			\begin{equation*}
				m :=  \frac{1}{\sum_{a} N_{f,a} } \sum_{a=1} ^S N_{f, a} m_a , \qquad  \lambda := t/ m R .
			\end{equation*}
			It is evident that all the cases studied until now are special cases of this generalized setting. For example, two sets carrying equal number of hypermultiplets as in Subsection \ref{sect:rhotwofamFI} correspond to $N_{f,1} = N_{f,2} \equiv N_f$ with $\epsilon (1) = -1 = - \epsilon (2)$.\par
			The generalized saddle point equation for the eigenvalue density in the decompactification limit is:
			\begin{equation}
			\label{eq:generalsaddlepointeq}
				\int_{- A} ^{B} \dd y \rho (y) \mathrm{sign} \left( x - y \right) - \frac{x}{m \lambda}  -  \sum_{a=1} ^S  \frac{ \chi _a }{2} \mathrm{sign} \left( x - \widetilde{m} _a \right)  = 0 ,
			\end{equation}
			where we introduced the Veneziano parameters $ \chi_a := \frac{ N_{ f,a} }{N} $ and called $\widetilde{m}_a = \epsilon (a) m_a + \zeta $ the masses shifted by the FI term. For simplicity in the forthcoming discussion, we arrange te masses into two sets: $\widetilde{m}^{-} _{a} < 0, a=1, \dots S^{-}$ and $\widetilde{m}^{+} _{b} > 0, b=1, \dots S^{+}$, with $S^{+} + S^{-} = S$, and relabel the shifted masses by increasing module:
			\begin{equation*}
				- \infty < \widetilde{m} ^{-} _{S^{-}} \le  \dots \le \widetilde{m} ^{-} _2 \le  \widetilde{m} ^{-} _1 \le 0 \le   \widetilde{m} ^{+} _1 \dots \le \widetilde{m}^{+} _{S^{+}} < + \infty .
			\end{equation*}
			We can figure out the phase structure at large $N$ in this generalized setting. For simplicity, we will also assume\footnote{This order holds automatically in the on-shell case.} $ | \widetilde{m}^{-} _{a} | < | \widetilde{m}^{+} _{a} |$; if not, the following analysis holds inverting the roles of $A$ and $B$.
			\begin{itemize}
				\item First a phase corresponding to small values of $A,B$, will appear, with uniform distribution $\frac{1}{2m \lambda}$. All resonances fall out of the domain and do not play any role.
				\item When $\lambda$ is such that $A$ approaches $\widetilde{m}^{-} _1$, there is a phase transition. The first negative resonance appears, placed at the negative boundary $-A$ of the domain.
				\item After that, depending on the $\chi_a$s, one of the following will happen: 
					\begin{enumerate}[(a)]
						\item $A$ increases while $B$ is kept below the resonance value. The first resonance now is placed in the interior of the domain, with coefficient $\chi_{1} ^{-} /2$.
						\item $A$ is kept at its value $\widetilde{m}^{-} _1$ and $B$ approaches $\widetilde{m}^{+} _1$. The eigenvalue density involves two Dirac deltas placed at its boundaries.
					\end{enumerate}
				\item If in the previous phase scenario (b) was realized, then $A$ moves away from its first critical value, with $B$ fixed at the first positive resonance value. If it was scenario (a) instead, then one of the following happens, depending on the $\chi_a$: either $B$ approaches the resonance value, so the first positive resonance appears at the boundary, or a second negative resonance is reached, cropping up at the negative boundary, due to the increasing value of $A$. If $| \widetilde{m}^{-} _a | < | \widetilde{m}^{+} _a | < | \widetilde{m}^{-} _{a+1} |$, this latter case is not allowed.
				\item This goes on until $A$ and $B$ overcome the values of all effective masses $\widetilde{m}^{\pm} _a$.
				\item Eventually, there is a phase with symmetric domain, $A=B > \widetilde{m}^{+} _{S^{+}} = \max \widetilde{m}_a $. In this phase the eigenvalue density yields a full sum of Dirac deltas, one at each $\widetilde{m}_a$, each one with coefficient $\chi_a /2$.
			\end{itemize}\par
			From the latter analysis we deduce that for $S$ mass scales we have, in general, $2S+1$ phases, with transitions determined by the boundaries of the support of $\rho$ hitting a resonance or overpassing it. However, in the generic setting a full concrete study is not feasible. The case of two different sets of hypermultiplets $\nu_{-} \ne \nu_{+}$ is reported in Appendix \ref{app:tablerho} as a further example.

		\subsection{Comment on complex masses}
		\label{sec:complexmasses}
			In the present work, the masses and the (analytically continued) FI coupling are taken to be real. Those parameters are introduced at field theoretical level by coupling, respectively, the chiral and vector multiplet to suitable Abelian background vector multiplets. Fields of those extra multiplets are non-dynamical and, after localization, their background configurations appear in the action as parameters. As explained in \cite{Jafferis:2010un}, masses introduced in this way can be given an additional imaginary part preserving rigid supersymmetry, with resulting complex mass parameters $m_{\mathbb{C},a} = m_a + i t_a /R$, where the $t_a$s parameterize the choice of R-symmetry (one $t_a$ for each mass scale $m_a$). The effect of such complexification is sub-leading in the large radius limit, hence the results discussed above hold also for the present case\footnote{Notice that, even though one sends $t_a\to \infty$ together with $R$, one would still obtain $\mathrm{sign} (x-m)$ in the saddle point equation, and the dependence on the imaginary part is dropped. We also note that, for the case of a single mass scale, one can obtain an arbitrary complex mass leaving the FI coupling without analytic continuation.}.\par
			More general complex mass assignments can be implemented introducing quadratic superpotentials for the hypermultiplets\footnote{See, for example, \cite{Willett:2016adv} for a review of the construction of $3d$ $\mathcal{N}=2$ gauge theories.}. Nonetheless, superpotential terms do not change the partition function on $\mathbb{S}^3$, since they vanish at the localization locus and do not contribute to the one-loop determinants \cite{Jafferis:2010un}. Thus, after localization, complex masses are allowed only if of the form $m_{\mathbb{C},a} = m_a + i t_a/R$. The resulting matrix model is more involved and contains hyperbolic functions of complex parameters. Nevertheless, while working in the decompactification limit, our results extend to the case of complex masses as well. At finite radius, however, the system of saddle point equations acquires imaginary part and becomes inconsistent, unless one allows the eigenvalues to be complex. To do so, we have to pass from an Hermitian matrix model, with eigenvalues distributed along the real line, to an holomorphic matrix model \cite{Lazaroiu:2003vh}. We comment on the formulation and difficulties in Appendix \ref{app:holomorphicMM}.

	\section{Antisymmetric Wilson loop with only mass deformation}
	\label{sect:WLmassL}
		
		In this second part of the work, we study the Wilson loop in the antisymmetric representation of rank $k$ in the decompactification limit, corresponding to different eigenvalue densities obtained in previous sections. We start with the case of two sets of $N_f$ hypermultiplets. The FI coupling is set to $\zeta =0$ and the Veneziano parameter $\chi = \frac{N_f}{N}$ is kept fixed at large $N$.\par
		Following \cite{Hartnoll:2006} we see that, at large $N$, the circular Wilson loop in antisymmetric representation of rank $k$ is given by:
		\begin{equation}
		\label{eq:WLantisymexpr}
			\langle W_k \rangle = \oint_{\alpha} \frac{ \dd w }{2 \pi \I } \frac{1}{ w ^{1+k}} \exp \left\{ N \int_{-A} ^B \dd x \rho (x) \log \left( 1 + w e^{x} \right) \right\} ,
		\end{equation}
		where $\rho$ is the eigenvalue density at large $N$, supported in $\left[ -A, B \right]$, which in the present case \cite{Barranco:2014} is to be read off from Section \ref{sect:BarrancoRusso}. The integration contour $\alpha$ is a circle in the complex plane around $w=0$. We are interested in the decompactification limit with $k$ comparable to $N$, keeping
		\begin{equation}
			f := \frac{ k}{N}
		\end{equation}
		fixed when sending $N \to \infty$. We change variable $w= e^{AR z}$ and, for the integral in the exponential, we set $x= - A y$. According to the discussion of the first part, $A$ has mass dimensions, hence $ARz$ and $y$ are adimensional. The change of variable in the complex plane maps the integration contour $\alpha$ into a circle $\gamma$ wrapping around the cylinder of radius $\frac{1}{AR}$. Shrinking the original contour $\alpha$ around the origin sends the circle around the cylinder to $\gamma \to \left\{ - \infty \right\} \times \cs^1 _{\frac{1}{AR}}$, while, on the other hand, we can adjust the original path $\alpha$ to get $\re z$ around $0$. The expression for the Wilson loop then reads
		\begin{equation}
		\label{eq:WLgeneralformbeforeSP}
			\langle W (f) \rangle = \frac{ AR }{2 \pi \I } \oint_{\gamma} \dd z \exp \left\{ A R N \left[ \int_{-B/A} ^{1} \frac{ \dd y }{R} \rho \left( - A y \right) \log  \left( 1 + e^{- AR \left( y - z \right) } \right) - f z \right] \right\} .
		\end{equation}
		The leading order contribution in the decompactification limit is obtained from the general saddle point equation:
		\begin{equation}
		\label{eq:saddlepointeqgen}
			\int_{-B/A} ^{1} A \dd y \rho \left( - A y \right) \frac{1}{1 + e^{ AR \left( y - z \right) }} - f = 0 .
		\end{equation}
		A glimpse into the structure of the saddle point equation suggests that no solution is expected at $\re z > 1$, as in this case $AR \left( y - z \right)<0$ and 
		\begin{equation*}
			\frac{1}{1 + e^{ AR \left( y - z \right) }} \xrightarrow{R \to \infty } 1 ,
		\end{equation*}
		reducing the saddle point equation to
		\begin{equation*}
			f = \int_{-A} ^B \dd x \rho (x) \equiv 1 ,
		\end{equation*}
		with last identity due to normalization. Analogously, for $\re z < - \frac{B}{A}$ the factor $\frac{1}{1 + e^{ AR \left( y - z \right) }}$ suppresses the integral and the equation is reduced to the trivial one $f=0$. A more detailed discussion is provided in Appendix \ref{app:noregSPWL}.
		
		\subsection{Saddle point contribution to the Wilson loop}
		
			To proceed further, we introduce the Barranco--Russo eigenvalue density \cite{Barranco:2014} of Section \ref{sect:BarrancoRusso}:
			\begin{equation}
				\rho (x) = \frac{1}{2 m \lambda} + \cp \left[ \delta \left( x + m \right) + \delta \left( x - m \right) \right] , \qquad x \in \left[ - A, A \right] ,
			\end{equation}
			which has symmetric domain $A=B$ in each phase, and with coefficients $\cI= 0, \cII = \frac{ \lambda - 1 }{2 \lambda} , \cIII= \frac{\chi}{2}$ depending on the phase. The symmetry of the domain implements the reflection symmetry
			\begin{equation}
				z \mapsto - z , \quad f \mapsto 1 - f ,
			\end{equation}
			which will be manifest in the solution. After a suitable change of variables\footnote{We will drop the $\prime$ from now on.} $y ^{\prime} = AR \left( y - z \right)$, we split the integral in the saddle point equation into three contributions:
			\begin{equation}
			\label{eq:saddlepoint3integrals}
				\frac{1}{2m R \lambda} \mathcal{I}_1 + \cp \left[ \mathcal{I}_2 + \mathcal{I}_3 \right] = f ,
			\end{equation}
			with
			\begin{equation}
			\begin{aligned}
				\mathcal{I}_1 & :=  \int_{- AR (1+z) } ^{ AR (1-z) } \dd y \frac{1}{1 + e^{y} } \\
						& = \left[ y - \log \left( 1 + e^y \right) \right]_{- AR (1+z) } ^{ AR (1-z) } = 2 AR  + \log \left( \frac{1 + e^{- AR (1+z) } }{1 + e^{AR (1-z)} } \right) ; \\
				\mathcal{I}_2 & := \int_{- AR (1+z) } ^{ AR (1-z) } \dd y \frac{1}{1 + e^{y} } \delta \left( y + AR z + Rm \right) = \frac{1}{1 + e^{- AR (z + \frac{m}{A}) } } ; \\
				\mathcal{I}_3 & := \int_{- AR (1+z) } ^{ AR (1-z) } \dd y \frac{1}{1 + e^{y} } \delta \left( y + AR z - Rm \right) = \frac{1}{1 + e^{- AR (z - \frac{m}{A}) } } .
			\end{aligned}
			\end{equation}\par
			We know from \cite{Hartnoll:2006} that the general sadle point equation admits solution for $\im z = 0 \ \mathrm{mod} \ 2\pi$, besides we can manipulate conveniently the original contour $\alpha$ so that the circle $\gamma$ wraps the cylinder passing through $-1 \le \re z \le 1$. We can actually do that because the branch cut of the logarithm in the expression \eqref{eq:WLantisymexpr} for $\langle W (f) \rangle$ lies at
			\begin{equation*}
				1 + e^{-AR \left( y - z \right) }= 0 \quad \Longrightarrow \quad \im z = \frac{\pi}{AR} (2n + 1) , \ n \in \Z .
			\end{equation*}\par
			In what follows we will solve the saddle point equation \eqref{eq:saddlepoint3integrals} for the relevant region $- 1 < \re z < 1$, while the other regions are discussed in Appendix \ref{app:noregSPWL}.

			\subsubsection{$-1 < \re z < 1$}
				Motivated by the general arguments above, we focus on the only relevant region $\im z = 0$ and $-1 < \re z < 1$, where the solution to the saddle point equation \eqref{eq:saddlepoint3integrals} has nontrivial structure. For $z$ in such range,
				\begin{equation*}
					1 + e^{- A R (1+z) } \to 1 , \quad 1 + e^{A R (1-z) } \to e^{AR (1-z)} ,
				\end{equation*}
				and so
				\begin{equation*}
					\mathcal{I}_1 \to 2 AR + \log \left( e^{-AR (1-z)} \right) = AR (1+z) .
				\end{equation*}
				For what concerns the other two integral contributions, we get:
				\begin{equation*}
				\begin{aligned}
					\mathcal{I}_2 & \to \begin{cases} 0 , \quad z< - \frac{m}{A} \\ 1,  \quad z>  - \frac{m}{A} \end{cases} = \theta \left( z + \frac{m}{A} \right) , \\
					\mathcal{I}_3 & \to \begin{cases} 0 , \quad z< \frac{m}{A} \\ 1,  \quad z>  \frac{m}{A} \end{cases} = \theta \left( z - \frac{m}{A} \right) ,
				\end{aligned}
				\end{equation*}
				where $\theta \left( \cdot \right)$ stands for the Heaviside step function. The correct value of the integrals at $z = \pm \frac{m}{A}$ is recovered assigning $\theta (0) = \frac{1}{2}$. The saddle point equation then becomes:
				\begin{equation*}
					\frac{A}{2 m \lambda} (1+z) + \cp \left[ \theta \left( z + \frac{m}{A} \right) + \theta \left( z - \frac{m}{A} \right) \right] = f ,
				\end{equation*}
				with formal solution:
				\begin{equation}
					1 + z = \frac{2 m \lambda}{A} \left( f - \cp  \left[ \theta \left( z + \frac{m}{A} \right) + \theta \left( z - \frac{m}{A} \right) \right] \right) .
				\end{equation}
				Depending on which phase we are considering, we can explicitly insert the values of $A$ and $\cp$ and obtain the saddle point.
				\begin{enumerate}[(I)]
					\item In phase I $A=m \lambda$ and $\cI=0$, from which the saddle point is simply
						\begin{equation}
							z = 2 f -1 ,
						\end{equation}
						which clearly respects the reflection symmetry $z \mapsto - z, f \mapsto 1-f$.
					\item In the second phase $A=m$ and $\cII = \frac{ \lambda - 1 }{2 \lambda}$. In the region considered, the first $\theta$ function is identically one whilst the second never contributes, hence the solution is:
						\begin{equation}
							z= \lambda \left( 2 f -1 \right) ,
						\end{equation}
						where again we can explicitly prove that the reflection symmetry holds.
					\item The situation in the third phase is more involved, due to $\frac{m}{A}= \frac{1}{\lambda \left( 1 - \chi \right)} < 1$, and the saddle point equation takes three different forms depending on the value of $z$. We also recall that $\cIII= \frac{\chi}{2}$.
						\begin{enumerate}[(i)]
							\item $-1 < \re z < - \frac{1}{\lambda \left( 1 - \chi \right)}$, where both $\theta$ functions vanish and the solution of the saddle point equation is:
								\begin{equation}
									z= \frac{ 2 f -1 + \chi }{1 - \chi } .
								\end{equation}
								The consistency condition for this solution is:
								\begin{equation*}
									\frac{ 2 f -1 + \chi }{1 - \chi } < - \frac{1}{\lambda \left( 1 - \chi \right)}  \quad \Longrightarrow \quad f < \frac{ \lambda - 1 }{2 \lambda} - \frac{\chi}{2} .
								\end{equation*}
							\item $ - \frac{1}{\lambda \left( 1 - \chi \right)} < \re z <  \frac{1}{\lambda \left( 1 - \chi \right)}$, in which case the first $\theta$ function is one and the second vanished. The solution is given by:
								\begin{equation}
									z = \frac{ 2 f - 1 }{1 - \chi}
								\end{equation}
								which is consistent as long as
								\begin{equation*}
									 - \frac{1}{\lambda \left( 1 - \chi \right)} < \frac{ 2 f - 1 }{1 - \chi} <  \frac{1}{\lambda \left( 1 - \chi \right)}
								\end{equation*}
								is satisfied, corresponding to:
								\begin{equation}
									\frac{ \lambda - 1}{2 \lambda} < f < \frac{ \lambda - 1}{2 \lambda} .
								\end{equation}
							\item Eventually for $\re z > \frac{1}{\lambda \left( 1 - \chi \right)} $ both $\theta$ functions have unit value and the solution is:
								\begin{equation}
									z = \frac{ 2 f -1 - \chi }{1 - \chi} ,
								\end{equation}
								with consistency condition
								\begin{equation}
									f > \frac{\lambda + 1 }{2 \lambda} + \frac{ \chi}{2} .
								\end{equation}
						\end{enumerate}
						We notice that also in phase III the reflection symmetry is satisfied, but only in the restricted region $ - \frac{1}{\lambda \left( 1 - \chi \right)} < \re z <  \frac{1}{\lambda \left( 1 - \chi \right)}$. Furthermore, we mention the peculiar fact that for certain values of $f$ as a function of $\lambda$ no solution is available. As this phases extends to arbitrary values of $\lambda > \frac{1}{1-\chi}$, only three distinct values of $f$ (one for each sub-case) allow existence of a solution for all $\lambda$, which are:
						\begin{equation*}
							f= \frac{1 - \chi }{2} , \ \frac{1}{2} , \ \frac{ 1 + \chi }{2} .
						\end{equation*}
						For all those three values, the saddle point is $z=0$, consistent with the fact that, as $\lambda$ increases, the left (respectively right) region where we looked for a saddle point extends toward $z=0$, while the middle region shrinks and eventually collapses onto $z=0$ at $\lambda \to \infty$. Remarkably, this special point is the fixed point of the reflection symmetry.
				\end{enumerate}
				It is straightforward to see that the solutions are continuously connected at phase transitions. At the transition $\lambda =\frac{1}{1 - \chi} $ the range of validity of the middle solution for phase III stretches over all possible values $0\le f \le 1$, and indeed it is the case for which the solution coincides with the one from phase II at transition value. Moreover, from the general discussion of Section \ref{sect:BarrancoRusso} we know that at $\chi=0$ phase II collapses onto the point $\lambda=1$ and phase I and III are joined together: consistently the solutions in each case coincide at $\chi =0$.

		\subsection{Evaluation of the Wilson loop}
			At this point we have all the ingredients to evaluate the Wilson loop in the antisymmetric representation of rank $k=fN$ in the decompactification limit. First of all we evaluate the contribution of the saddle point, which is given by:
			\begin{equation*}
			\begin{aligned}
					S(z)= & \int_{-1} ^1 \frac{\dd y }{R} \rho \left( - A y \right) \log \left( 1 + e^{-A R \left( y - z \right)} \right) - f z = \\
					& \int_{-1} ^1 \dd y \left( \frac{1}{2 m R \lambda} + \frac{ \cp }{R} \left[ \delta \left( - A y + m \right) + \delta \left( A y + m \right) \right] \right) \log \left( 1 + e^{-A R \left( y - z \right)} \right) - f z .
			\end{aligned}
			\end{equation*}
			Taking the limit in the region $-1 < \re z < 1$, which we have seen to be the only relevant one, the first integral is nonvanishing only for $-1 \le y \le z$, and we are left with:
			\begin{equation}
				S(z) \to \frac{A}{4 m \lambda}\left( 1 + z \right)^2 + \frac{\cp}{A R } \left[ \log \left(1 + e^{A R \left( z + \frac{m}{a} \right)} \right) + \log \left(1 + e^{A \left( z - \frac{m}{a} \right)} \right)  \right] - fz ,
			\end{equation}
			to be evaluated at $z$ solution of the saddle point equation. Recalling the expression \eqref{eq:WLgeneralformbeforeSP} for the Wilson loop, we have that in the decompactification limit it is given by the saddle point contribution, according to:
			\begin{equation}
			\label{eq:WLsaddlepointexpr}
				\langle W(f) \rangle ^{(0)} = \frac{AR}{2 \pi i} \oint_{\gamma} \dd z e^{A R N S(z)} = e^{ A R N S(z_s)} ,
			\end{equation}
			for $z_s$ satisfying the saddle point condition, where the second equality follows from the  definition of the curve $\gamma$.
			\begin{enumerate}[(I)]
				\item In the first phase we get:
					\begin{equation*}
						S(z_s) = \left[ \frac{(1+z)^2}{4} - fz \right]_{z= 2f -1} = f (1-f).
					\end{equation*}
					Together with the general formula \eqref{eq:WLsaddlepointexpr} for the Wilson loop we arrive to:
					\begin{equation}
						\langle W(f) \rangle ^{(0)} \vert_{\mathrm{I}} = e^{ m \lambda R N f (1-f) } .
					\end{equation}
				\item Approximating the logarithms at large radius and inserting the parameters from phase II we have:
					\begin{equation*}
						S(z_s) = \left[ \frac{ \left( 1 + z \right)^2}{4 \lambda} + \frac{ \lambda - 1 }{2 \lambda} \left( 1 + z \right) - f z \right]_{z=\lambda \left( 2 f - 1 \right)} = \lambda \left[ f \left( 1 - f \right) - \left( \frac{ \lambda - 1 }{2\lambda} \right)^2 \right] .
					\end{equation*}
					THe Wilson loop is given by:
					\begin{equation}
						\langle W(f) \rangle ^{(0)} \vert_{\mathrm{II}} = e^{ m \lambda R N \left[ f (1-f)  - \left( \frac{ \lambda - 1 }{2\lambda} \right)^2 \right] } .
					\end{equation}
				\item For phase III we can do the same approximation for the logarithms, but now we have to split the study according to the different subregions for the solution.
					\begin{enumerate}[(i)]
						\item For $f < \frac{ \lambda - 1 }{2 \lambda} - \frac{\chi}{2}$ the contribution coming from the saddle point is:
							\begin{equation*}
								S(z_s) = \left[ \frac{ \left( 1 - \chi \right) \left( 1 + z \right)^2 }{4} - f z \right]_{z = \frac{ 2 f - 1 + \chi }{1 - \chi }} = \frac{ f \left( 1 - \chi - f \right) }{1 - \chi} ,
							\end{equation*}
							which, once inserted in \eqref{eq:WLsaddlepointexpr} gives:
							\begin{equation}
								\langle W(f) \rangle ^{(0)} \vert_{\mathrm{III}} = e^{ m \lambda R N f \left( 1 - \chi - f \right) } .
							\end{equation}
						\item In the intermediate region $\frac{ \lambda - 1 }{2 \lambda} < f < \frac{ \lambda + 1 }{2 \lambda}$ the saddle point contribution includes the approximation of the first logarithm, hence:
							\begin{equation*}
							\begin{aligned}
								S (z_s) & = \left[ \frac{ \left( 1 - \chi \right) \left( 1 + z \right)^2 }{4} + \frac{\chi}{2} \left( z + \frac{1}{\lambda \left( 1 - \chi \right)} \right)- f z \right]_{z = \frac{ 2 f - 1  }{1 - \chi }} \\
										& =  \frac{1}{1 - \chi} \left[ f \left( 1-f \right) - \chi \left( \frac{ \lambda - 1 }{2 \lambda } - \frac{\chi}{4} \right) \right] .
							\end{aligned}
							\end{equation*}
							We use it to evaluate the Wilson loop getting:
							\begin{equation}
								\langle W(f) \rangle ^{(0)} \vert_{\mathrm{III}} = e^{ m \lambda R N \left[ f \left( 1-f \right) - \chi \left( \frac{ \lambda - 1 }{2 \lambda } - \frac{\chi}{4} \right) \right] } .
							\end{equation}
						\item When the saddle point is placed above $\frac{1}{\lambda \left( 1 - \chi \right)}$, which corresponds to $f> \frac{ \lambda + 1 }{2 \lambda} + \frac{\chi}{2}$, both logarithms contribute, leading to:
							\begin{equation*}
							\begin{aligned}
								S (z_s) & = \left[ \frac{ \left( 1 - \chi \right) \left( 1 + z \right)^2 }{4} + \frac{\chi}{2} \left\{ \left( z + \frac{1}{\lambda \left( 1 - \chi \right)} \right) + \left( z - \frac{1}{\lambda \left( 1 - \chi \right)} \right) \right\} - f z \right]_{z = \frac{ 2 f - 1 - \chi  }{1 - \chi }} \\
									& = \frac{ \left( f - \chi \right) \left( 1 -f \right) }{1 - \chi} .
							\end{aligned}
							\end{equation*}
							Substituting this latter expression into \eqref{eq:WLsaddlepointexpr} we obtain the Wilson loop:
							\begin{equation}
								\langle W(f) \rangle ^{(0)} \vert_{\mathrm{III}} = e^{ m \lambda R N  \left( f- \chi \right) \left( 1-f \right) } .
							\end{equation}
					\end{enumerate}
					We therefore have an expression for the Wilson loop in phase III for all values of $f$ that admit saddle point. We can write it in a compact way as:
					\begin{equation}
						\langle W(f) \rangle ^{(0)} \vert_{\mathrm{III}} = e^{ m \lambda R N \left[ f \left( 1-f \right) - \chi \beta_{\lambda} (f) \right]} ,
					\end{equation}
					where we have introduced
					\begin{equation}
					\label{eq:betashortcut}
						\beta_{\lambda} (f) = \begin{cases} f   \\   \frac{\lambda -1}{2 \lambda} - \frac{\chi}{4}  ,  \\  1 - f , \end{cases} \ \begin{matrix} 0 & \le f < & \frac{\lambda -1}{2 \lambda} - \frac{\chi}{2} , \\ \frac{\lambda -1}{2 \lambda} & < f < &  \frac{\lambda +1}{2 \lambda} , \\  \frac{\lambda +1}{2 \lambda} + \frac{\chi}{2} & < f \le & 1 , \end{matrix}
					\end{equation}
					whilst no solution to the saddle point equation exists for other values of $f$.
			\end{enumerate}
			Wilson loop is continuous at critical values $\lambda =1$ and $\lambda= \frac{1}{1 - \chi}$, where, for what concerns phase III, the middle region extends to the whole $-1 < \re z <1$ as $\lambda$ approaches the critical value, thus the expression to be used is (ii).
			
		\subsection{Wilson loop analysis}
			The Wilson loop satisfies a perimeter law analogous to the one of \cite{Anderson:2015,Russo:2017ngf}.\par
			The ``energy'' associated to the Wilson loop is defined as:
			\begin{equation}
				\mathcal{W} := - \frac{1}{R N} \log \langle W(f) \rangle ,
			\end{equation}
			The energy and its first derivative are continuous (Appendix \ref{app:WLfirstderiv}), but the second derivative presents finite discontinuities: the phase transition is of second order for the Wilson loop, as was the case for the four-dimensional model in \cite{Russo:2017ngf}.
			
		\subsubsection{The $m \to 0$ and $2 N_f \to 0$ limit }
		\label{sec:subsecspeciallimit1}
			We study now a limit where we first reduce to a super-conformal theory \cite{Gaiotto:2007qi} $m \to 0$ and then we send the number of hypermultiplets $2 N_f \to 0$. To be consistent with the decompactification limit, we have to turn off the mass term and at the same time increase $\lambda$, so that the scaling $t/R$ is kept fixed. In this way, and taking into account that for $\lambda$ big enough the system is in phase III, we have:
			\begin{equation}
				\langle W (f) \rangle ^{(0)} \xrightarrow{ \ m \to 0 \ } e^{t R \left[ f \left( 1 - f \right) - \chi \beta_{\infty} (f) \right] } \xrightarrow{ \ N_f \to 0 \ } e^{ t N f ( 1 - f)} .
			\end{equation}
			Such expression still behave according to a perimeter law, as $t$ scales as $R$ in the decompactification limit. Furthermore, from the discussion about the saddle point in phase III, we know that the solutions at $\lambda \to \infty$ only makes sense for any of the three values of $f \in \left\{ \frac{ 1 - \chi}{2} , \frac{1}{2} , \frac{1+\chi}{2} \right\}$, which for $N_{f} \to 0$ ($\chi \to 0$) collapse onto $f=\frac{1}{2}$. Note that, $\chi=0$ is the unique value for which a solution exists for all $f$, and also it extends the reflection symmetry to all three sub-cases of phase III. Moreover, the solution $f=\frac{1}{2}$ corresponds to the fixed point of such symmetry, and is exactly the value that maximizes the Wilson loop.\par
			Conversely, if we first set $\chi=0$ the constraint on $f$ is removed, and sending $m \to 0$ afterwards gives formally the same expression for the Wilson loop, but now the solution exists for all $0 \le f \le 1$.
			
		\subsubsection{Decoupling matter and pure Chern-Simons theory}
		\label{sec:subsecspeciallimit2}

			If we take the $m \to \infty $ limit, the matter hypermultiplets decouple and we are left with an expression in terms of the pure Chern--Simons matrix model. Then, for any representation $\mathcal{R}$ of $U(N)$, the Chern--Simons matrix model evaluation of the Wilson loop on $\cs^3$ gives \cite{Dolivet:2006ii}
			\begin{equation}
				\langle W( \mathcal{R}) \rangle = \dim_q (\mathcal{R}) q^{- \frac{1}{2} c_2 (\mathcal{R}) } ,
			\end{equation}
			with deformation parameter $q= e^{-g_s}$. The quantum dimension is defined as
			\begin{equation*}
				\dim_q (\mathcal{R}) = \prod_{1 \le i < j \le N } \frac{ \left[ \mathcal{R}_i - \mathcal{R}_j - i + j \right]_q }{\left[ -i + j \right]_q } ,
			\end{equation*}
			where $\mathcal{R}_i$ means the length of the $i$-th box of the Young diagram associated to $\mathcal{R}$, and $\left[ \cdot \right]_q$ denotes the $q$-number. Besides, $c_2 (\mathcal{R})$ is the quadratic Casimir of the representation $\mathcal{R}$:
			\begin{equation*}
				c_2 ( \mathcal{R} ) = \sum_{j=1} ^N \mathcal{R}_j \left( \mathcal{R}_j + N + 1 -2j \right) .
			\end{equation*}
			In our case, $\mathcal{R}$ is chosen to be the antisymmetric representation of rank $k$, with $f= \frac{k}{N}$ fixed at large $N$, which corresponds to
			\begin{equation*}
				\mathcal{R}_i = \begin{cases} 1 , \quad i \le k \\ 0, \quad i > k . \end{cases}
			\end{equation*}
			The formula for the quantum dimension reads:
			\begin{equation*}
			\begin{aligned}
				\dim_q (\mathcal{R}) & = \prod_{1 \le i \le k < j \le N } \frac{ \left[ 1 - i + j \right]_q }{\left[ -i + j \right]_q }  =  \prod_{l=1} ^{N-k} \prod_{m=0}^{k-1} \frac{ \left[ l + m +1 \right]_q }{ \left[ l + m  \right]_q } = \prod_{j=1} ^{N-k} \frac{ \left[ k+ j \right]_q }{ \left[ j \right]_q } \\
					& \approx \frac{  g_s ^{-N + k} \exp \left( \frac{g_s}{2} \sum_{j=1} ^{N-k} \left( k + j \right) \right) }{ g_s ^{-N +k} \exp \left( \frac{g_s}{2} \sum_{j=1} ^{N-k} j \right) }= \exp \left( \frac{ t N }{2} f (1 - f) \right)
			\end{aligned}
			\end{equation*}
		where for the second equality we changed variables to $j=k+l$, $i=k-m$, and in the last part we approximated the $q$-numbers at small $g_s$. We are considering the large $N$ limit, so most of the contributions are a product of ones. Note that the last approximation is good for the numerator at large $N$ but works bad for the first terms of the denominator, which are order 1, thus not diverging at large $N$. Nevertheless, a more careful insight using Stirling formula shows that terms we neglected are lower order in $N$. As for the part with the Casimir, it reads
		\begin{equation*}
			c_2 ( \mathcal{R} ) = \sum_{i=1} ^k  \left(  N +2 -2i \right) = k (N + 1 - k) \approx  N^2 f (1-f) ,
		\end{equation*}
		where again we are neglecting terms of lower order in $N$. Putting all together, we find
		\begin{equation}
			\langle W (f) \rangle \approx e^{ \frac{ t N }{2} f (1 - f) } q^{- \frac{N^2}{2} f(1-f) } = e^{t N f (1-f)} ,
		\end{equation}
		We recover our previous result in phase I under the identification $g_s N = t = m R \lambda$. We underline that, although we are in the regime $q \to 1$, we cannot directly replace the quantum dimension with usual dimension, since terms $\sim q^{-N} = e^{ g_s N} = e^{m R \lambda }$ are relevant in the decompactification limit.\par
		As a concluding remark we notice that is not surprising that results from pure Chern--Simons matrix model are to be compared with phase I of the present model: indeed, in the $m \to \infty$ limit, we get rid of the resonances and hence the system never leaves phase I.

	\section{$\mathcal{O} \left( 1 / R \right)$ corrections to the Wilson loop}
	\label{sect:oRcorrectionsWL}
		
		In this section we include $\oR$ corrections to the Wilson loop in the antisymmetric representation, with only mass deformation. We do that by using the complete expressions for $A$ and $\rho$, which are given in \cite{Barranco:2014} (Eq. (4.23) and Eq. (4.24)), and approximate it at large radius keeping track of $\oR$ terms. After tedious but straightforward calculations, where we use approximations such as:
		\begin{equation*}
			\log \cosh \frac{AR}{2} = \log \frac{e^{AR/2}}{2} \left( 1 + e^{-AR} \right) \approx \frac{AR}{2} - \log 2 + e^{-AR}  ,
		\end{equation*}
		we arrive to the equation
		\begin{equation}
		\label{eq:approxexprA}
			AR - \log 4 = m R \lambda \left( 1 - \chi \right) + t \chi  \frac{ e^{ \left( m - A \right)R /4 }  }{ \left( e^{ \left( m - A \right)R /2 } + e^{ - \left( m - A \right)R /2 } \right)^{1/2}}, 
		\end{equation}
		to determine the value of $A$. This latter expression can be further approximated, and then solved, depending on whether the control parameter $\lambda$ is such that $A<m$ or $A>m$. More specifically, if $A<m$ the last term in the equation goes to $1$, and if $A>m$ it is suppressed and goes to $0$. Denoting by $\lambda_{-}$ (respectively $\lambda_{+}$) the value such that $A<m$ for $\lambda < \lambda_{-}$ (respectively $A>m$ for $\lambda > \lambda_{+}$) we obtain:\begingroup
\renewcommand*{\arraystretch}{1.2}
		\begin{equation}
			A = \begin{cases} m \lambda + \frac{ \log 4 }{R} , \\  m \lambda \left( 1 - \chi \right) + \frac{ \log 4 }{R} , \end{cases} \ \begin{matrix} \ & \quad \lambda < & \lambda_{-} , \\  \ & \quad \lambda > & \lambda_{+} ,\end{matrix}
		\end{equation}\endgroup
		where we have dropped exponentially suppressed corrections and 
		\begin{equation*}
			\lambda_{-} = 1 + \oR , \quad \lambda_{+} = \frac{1}{1- \chi } + \oR .
		\end{equation*}
		For the moment we avoid the study of what happens for $\lambda_{-} < \lambda < \lambda_{+}$, corresponding to phase II, and delay it to Appendix \ref{app:phaseIIORcorrections}. We see that the three-phase structure persists and, in phases I and III, the $\oR$ correction received by $A$ is a constant shift, and all other terms are exponentially decaying. Therefore we expect the Wilson loop to maintain its qualitative behaviour.\par
		One could try to apply a similar procedure to $\rho$. Nevertheless one soon sees that no polynomial term in $A$ appears, and all corrections are exponentially decaying. Therefore, for what concerns phases I and III, the eigenvalue density is the same when we include $\oR$ corrections. The case of phase II deserves more attention, as argued in Appendix \ref{app:phaseIIORcorrections}.\par
		We recall that, using the same expression for $\rho$ as above, the saddle point equation is:
		\begin{equation*}
			\frac{1}{2 m R \lambda} \mathcal{I}_1  + \cp \left[ \mathcal{I}_2 + \mathcal{I}_3 \right] = f ,
		\end{equation*}
		with
		\begin{equation*}
		\begin{aligned}
			\mathcal{I}_1 & = 2 A R + \log \left( \frac{ 1 + e^{-AR (1+z)}}{1+ e^{AR(1-z)}}  \right) , \\
			\mathcal{I}_2 & = \frac{1}{1 + e^{- AR \left( z + \frac{m}{A} \right) }} , \\ 
			\mathcal{I}_2 & = \frac{1}{1 + e^{- AR \left( z - \frac{m}{A} \right) }} .
		\end{aligned}
		\end{equation*}
		Besides, the general contribution to the exponential of the Wilson loop is given by the expression:
		\begin{equation}
		\label{eq:1Rcorrectionsgensaddlecontrib}
			S(z) = \frac{ A }{ 4 m \lambda} \left( 1 + z \right)^2 + \frac{ \cp }{AR} \left[ \log \left(  1 + e^{AR \left( z + \frac{m}{A} \right) } \right) + \log \left( 1 + e^{AR \left( z - \frac{m}{A} \right) } \right)  \right] - f z .
		\end{equation}\par
		We replicate the procedure of the previous section, now using implemented expressions for $A$. We will assume henceforth $-1 < \re < < 1$, which, as we have seen, is the only relevant case.
		\begin{enumerate}[(I)]
			\item When $\lambda < \lambda_{-}$ the eigenvalue density has support with boundary $A= m \lambda + \frac{ \log 4 }{R}$. By means of standard approximation used so far we get:
				\begin{equation*}
					\frac{1}{2 m R \lambda} \mathcal{I}_1 \approx \frac{1}{2} \left( 1 + \frac{ \log 4 }{m R \lambda } \right) \left( 1 + z \right) ,
				\end{equation*}
				which, together with $\cI = 0$, provides the saddle point:
				\begin{equation}
					1 + z = 2 f \left( 1 + \frac{ \log 4 }{m R \lambda } \right)^{-1} .
				\end{equation}
				The saddle point contribution to the Wilson loop in this case is then:
				\begin{equation*}
				\begin{aligned}
					S(z_s) & = \frac{1}{4} \left( 1 + \frac{ \log 4 }{m R \lambda } \right) \left[ 2 f \left( 1 + \frac{ \log 4 }{m R \lambda } \right)^{-1} \right]^2 - f \left[ 2 f \left( 1 + \frac{ \log 4 }{m R \lambda } \right)^{-1} - 1 \right] \\
						& = f \left( 1 - f  \left( 1 + \frac{ \log 4 }{m R \lambda } \right)^{-1} \right) ,
				\end{aligned}
				\end{equation*}
				form which we obtain the first order expression to the Wilson loop in the decompactification limit
				\begin{equation}
				\begin{aligned}
					\langle W (f) \rangle ^{(I)} \vert_{\mathrm{I}} & = \exp \left\{ m R \lambda \left( 1 + \frac{ \log 4 }{m R \lambda } \right) N f \left( 1 - f  \left( 1 + \frac{ \log 4 }{m R \lambda } \right)^{-1} \right)  \right\} \\
						& = e^{  m R \lambda N f \left(1-f \right) + N f \log 4 }.
				\end{aligned}
				\end{equation}
				We notice that $\langle W (f) \rangle ^{(I)} \vert_{\mathrm{I}} = 2^{2Nf } \langle W (f) \rangle ^{(0)} \vert_{\mathrm{I}} $, with the $\oR$ correction not respecting the perimeter law and scaling as $k= Nf$.
			\item Phase II is determined by the requirement that $A$ is comparable with $m$, in the sense that we can neglect their difference in the decompactification limit. Introducing $\oR$ corrections moves $A$ away from the exact value $A=m$. We skip any more detailed discussion here, and the interested reader may refer to Appendix \ref{app:phaseIIORcorrections}.
			\item When $\lambda > \lambda_{+}$ we enter the third phase, when both resonances are inside the domain of the eigenvalue density. Here, $A= m \lambda \left( 1 - \chi \right)  + \frac{ \log 4 }{R}$, so that
				\begin{equation*}
					\frac{m}{A} = \frac{1}{\lambda \left( 1 - \chi \right) } \left( 1 + \frac{ \log 4 }{m R \lambda \left( 1 - \chi \right)} \right)^{-1} .
				\end{equation*}
				The integrals in the saddle point equation can be approximated to:
				\begin{equation*}
				\begin{aligned}
					& \frac{1}{2 m R \lambda} \mathcal{I}_1  = \frac{ 1 - \chi}{2} \left( 1 + \frac{ \log 4 }{m R \lambda \left( 1 - \chi \right)} \right) \left( 1 + z \right) , \\
					& \mathcal{I}_2  = \theta \left( z + \frac{1}{\lambda \left( 1 - \chi \right) } \left( 1 + \frac{ \log 4 }{m R \lambda \left( 1 - \chi \right)} \right)^{-1} \right) , \\
					& \mathcal{I}_3  = \theta \left( z - \frac{1}{\lambda \left( 1 - \chi \right) } \left( 1 + \frac{ \log 4 }{m R \lambda \left( 1 - \chi \right)} \right)^{- 1} \right) ,
				\end{aligned}
				\end{equation*}
				where $\theta$ is the Heaviside step function. We notice that, for what concerns $\mathcal{I}_2$ and $\mathcal{I}_3$, the effect of the correction is just to move the discontinuity of $\theta$. The saddle point equation is:
				\begin{equation}
					\frac{ 1 - \chi}{2} \left( 1 + \frac{ \log 4 }{m R \lambda \left( 1 - \chi \right)} \right) \left( 1 + z \right) + \frac{ \chi }{2} \left[ \theta \left( z + \frac{m}{A} \right) +  \theta \left( z - \frac{m}{A} \right) \right] = f .
				\end{equation}
				As usual, we split the study in three sub-cases, depending on whether none, one or both $\theta$ functions contribute.
				\begin{enumerate}[(i)]
					\item For $\re z < - \frac{m}{A}$ the saddle point is given by
						\begin{equation}
							1 + z = \frac{2 f }{1 - \chi } \left( 1 + \frac{ \log 4 }{m R \lambda \left( 1 - \chi \right)} \right)^{-1} ,
						\end{equation}
						with consistency condition
						\begin{equation*}
							z = \frac{2 f }{1 - \chi } \left( 1 + \frac{ \log 4 }{m R \lambda \left( 1 - \chi \right)} \right)^{-1} - 1 < - \frac{1}{\lambda \left( 1 - \chi \right) } \left( 1 + \frac{ \log 4 }{m R \lambda \left( 1 - \chi \right)} \right) ,
						\end{equation*}
						which is satisfied for
						\begin{equation}
							f < \frac{ \lambda - 1 }{2 \lambda} - \frac{\chi}{2} + \frac{\log 4 }{2 m R \lambda} .
						\end{equation}
					\item In the region $ - \frac{m}{A} < \re z < \frac{m}{A}$ the first $\theta$ kicks in, and the saddle point is:
						\begin{equation}
							1 + z = \frac{2 f - \chi }{1 - \chi } \left( 1 + \frac{ \log 4 }{m R \lambda \left( 1 - \chi \right)} \right)^{-1} .
						\end{equation}
						Consistency condition holds for
						\begin{equation}
							 \frac{ \lambda - 1 }{2 \lambda}  + \frac{\log 4 }{2 m R \lambda} < f < \frac{ \lambda + 1 }{2 \lambda}  + \frac{\log 4 }{2 m R \lambda} .
						\end{equation}
					\item Eventually for $\re z > \frac{m}{A}$ both resonances play a role and the saddle point here is given by:
						\begin{equation}
							1 + z = \frac{2 \left( f - \chi \right)}{1 - \chi } \left( 1 + \frac{ \log 4 }{m R \lambda \left( 1 - \chi \right)} \right)^{-1} .
						\end{equation}
						The saddle point is placed in this third region for
						\begin{equation}
							f > \frac{ \lambda + 1 }{2 \lambda} + \frac{\chi}{2} + \frac{\log 4 }{2 m R \lambda} .
						\end{equation}
				\end{enumerate}
				Inserting those results in \eqref{eq:1Rcorrectionsgensaddlecontrib} we obtain the saddle point contribution at $\oR$. After calculations very similar to the ones already done we arrive to:
				\begin{equation}
					\langle W (f) \rangle ^{(I)} \vert_{\mathrm{III}} = e^{ m R \lambda N  \left[ f \left( 1 - f \right) - \chi \beta_{\lambda} ^{(L)} (f) + \frac{ \log 4 }{m R \lambda} \beta_{\lambda} ^{(NL)} (f) \right] } ,
				\end{equation}
				where $\beta_{\lambda} ^{(L)}$ in the leading order of the $\beta_{\lambda}$ function of Section \ref{sect:WLmassL}, given in \eqref{eq:betashortcut}, and its next-to-leading order correction is:\begingroup
\renewcommand*{\arraystretch}{1}
				\begin{equation}
					\beta_{\lambda} ^{(NL)} (f) = \begin{cases} f ,  \\ f - \frac{ \chi }{2} ,  \\ f - \chi , \end{cases} \ \begin{matrix} \ & \quad f < & \frac{ \lambda - 1}{2 \lambda} - \frac{ \chi}{2} + \frac{ \log 2 }{ m R \lambda  } ,  \\ \frac{ \lambda - 1}{2 \lambda} - \frac{ \chi}{2} + \frac{ \log 2 }{m R \lambda  } & < f  < & \frac{ \lambda + 1}{2 \lambda} + \frac{ \chi}{2} + \frac{ \log 2 }{m R \lambda  } , \\ \ & \quad f > & \frac{ \lambda + 1}{2 \lambda} + \frac{ \chi}{2} + \frac{ \log 2 }{m R \lambda  } . \end{matrix}
				\end{equation}\endgroup
				We again highlight the behaviour 
				\begin{equation*}
					\langle W (f) \rangle ^{(I)} \vert_{\mathrm{III}} = \langle W (f) \rangle ^{(0)} \vert_{\mathrm{III}} 2^{ 2 N \beta_{\lambda} ^{(NL)} (f) } ,
				\end{equation*}
				with the $\oR$ correction factorized which breaks the perimeter law, introducing a scaling with $k$.
		\end{enumerate}\par
		\medskip
		As expected by general analysis, $\oR$ corrections do not affect the phase structure, as they only introduce a $\lambda$-independent shift in $A$ and do not change the eigenvalue density. Consistently, the Wilson loop is modified by a shift in the exponential that depends not on $R$ nor $\lambda$.

	\section{Wilson loop with insertion of FI term}
	
		According to the features exposed at the beginning of Section \ref{sect:WLmassL}, the method developed in \cite{Hartnoll:2006} allows to obtain the Wilson loop in the antisymmetric representation of rank $k$ through \eqref{eq:WLgeneralformbeforeSP}:
		\begin{equation*}
			\langle W (f) \rangle = \frac{ AR }{2 \pi \I } \oint_{\gamma} \dd z \exp \left\{ A R N \left[ \int_{-B/A} ^{1} \frac{ \dd y }{R} \rho \left( - A y \right) \log  \left( 1 + e^{- AR \left( y - z \right) } \right) - f z \right] \right\} ,
		\end{equation*}
		where we recall that $f:= \frac{k }{N}$ is kept finite at large $N$, and $\gamma$ is a circle of radius $\frac{1}{AR}$. The saddle point equation for a general eigenvalue density $\rho$ is \eqref{eq:saddlepointeqgen}:
		\begin{equation*}
			\int_{-B/A} ^{1} A \dd y \rho \left( - A y \right) \frac{1}{1 + e^{ AR \left( y - z \right) }} - f = 0 .
		\end{equation*}
		From analysis of the integral term (see also Appendix \ref{app:noregSPWL}), we learn that we have to look for saddle points in the region $- \frac{ B}{A} < \re z < 1$.\par
		When the model considered includes the Fayet--Iliopoulos term, we have to plug in any of the distributions from Section \ref{sect:FItermdistrib}, which do not respect the reflection symmetry. When two mass scales are present, the general expression for the eigenvalue density is given by Eq. \eqref{eq:rhogenP}:
		\begin{equation*}
			\rho (x) = \frac{1}{2 m \lambda} + \cp ^{-} \delta \left( x + m_{-} \right) + \cp ^{+} \delta \left( x - m_{+} \right)  , \qquad x \in \left[ - A , B \right] ,
		\end{equation*}
		with coefficients $\cp ^{\pm}$ and boundaries $A,B,$ of the support depending on the phase. Using this general expression the saddle point equation can be rewritten as:
		\begin{equation}
			\frac{1 }{2 m R \lambda} \mathcal{I}_1 + \cp^{-} \mathcal{I}_2 +\cp^{+} \mathcal{I}_3 = f ,
		\end{equation}
		with integrals
		\begin{equation*}
		\begin{aligned}
			\mathcal{I}_1 & = \int_{- BR \left( 1 + z \right) } ^{AR \left( 1 - z \right)} \dd y \frac{1}{1+ e^{y} } = \left( A+B \right) R + \left( A - B \right)R z + \log \left( \frac{ 1 + e^{- AR \left( z + \frac{B}{A} \right) } }{1 + e^{AR \left(1 - z \right)}} \right) , \\
			\mathcal{I}_2 & = \frac{1}{1 + e^{- AR \left( z - \frac{m_{-}}{A} \right) }} , \\
			\mathcal{I}_3 & = \frac{1}{1 + e^{- AR \left( z + \frac{m_{+}}{A} \right) }} ,
		\end{aligned}
		\end{equation*}
		with solution for $\mathcal{I}_2$ holding for $A \ge m_{-}$ or vanishing otherwise, and solution for $\mathcal{I}_3$ holding for $B \ge m_{+}$ or vanishing otherwise. As in the $\zeta=0$ case, the logarithm $\mathcal{I}_1$ can be approximated at large radius, in the region of interest $- \frac{B}{A} < \re z < 1$, by $- AR (1-z)$, so that:
		\begin{equation}
			\mathcal{I}_1 = BR + \left( 2A - B \right)R z .
		\end{equation}
		For what concerns $\mathcal{I}_2$ and $\mathcal{I}_3$, in the decompactification limit they are either suppressed and vanish or converge to $1$:
		\begin{equation}
		\begin{aligned}
			\mathcal{I}_2 = \theta \left( z - \frac{m_{-}}{A} \right) , \\
			\mathcal{I}_3 = \theta \left( z + \frac{m_{+}}{A} \right) ,
		\end{aligned}
		\end{equation}
		with $\theta$ the Heaviside function. Therefore, putting all together we arrive to the formal solution to the saddle point equation:
		\begin{equation}
		\label{eq:gensaddlepointsolFI}
			z= \frac{ 2 m \lambda}{2 A - B } \left( f - \left[ \cp ^{-} \theta  \left( z - \frac{m_{-}}{A} \right) + \cp^{+}  \theta \left( z + \frac{m_{+}}{A} \right) \right] \right) - \frac{B}{2A - B} .
		\end{equation}
		From Section \ref{sect:FItermdistrib} we learned that five different phases arise in the decompactification limit: from the third region on, the pursuit of the saddle point splits into sub-regions, respectively for none, one or both nonvanishing $\theta$ functions.
		
		\subsection{Antisymmetric Wilson loop for $\zeta < m_1$ and two mass scales}
			We now provide explicit evaluation of the Wilson loop in the case of two sets with equal number of massive hypermultiplets, which in the above formalism means $\nu_{-} = \nu_{+} =1$, and messes $-m_1, m_2$. We moreover consider $0 \le \zeta \le m_1$, so that one resonance is at negative and one at positive eigenvalues. We work not necessarily on-shell, but henceforth we will assume $m_{-} < m_{+}$. The phase-by-phase expression for the density $\rho$ can be retrieved in Subsection \ref{sect:rhotwofamFI}.

		\subsubsection{Saddle point for $\zeta < m_1$}
			We now calculate explicitly the saddle point in each phase, using results from Subsection \ref{sect:rhotwofamFI} and expression \eqref{eq:gensaddlepointsolFI}. In the next subsection we will use the results that follow to evaluate the Wilson loop.\par
			\begin{enumerate}[(I)]
				\item In the first phase the saddle point is $z = 2f - 1 $.
				\item The saddle point in the second phase, from \eqref{eq:gensaddlepointsolFI}, is:
					\begin{equation*}
						z= \frac{ m \lambda }{ 2 m_{-} - m \lambda} \left( 2 f - 1 \right) .
					\end{equation*}
				\item We recall that phase III can take two different forms, corresponding to $\chi$ above or below a certain critical value.
					\begin{enumerate}[(a)]
						\item If $\chi < 1 - \frac{m_{-}}{m_{+}}$ phase III holds for $\frac{m_{-}}{m \left( 1 - \chi \right)} < \lambda < \frac{m_{+}}{m}$. To find the saddle point, we fist have to look for solutions in different sub-regions and check the consistency of each solution.
							\begin{enumerate}[(i)]
								\item For $z < \frac{ m_{-}}{A}$ the saddle point is
								\begin{equation*}
									z = \frac{ 2 f - 1}{1 - 2 \chi} ,
								\end{equation*}
								consistent for values of the parameter
									\begin{equation*}
										f< \frac{1}{2} \left( 1+ \frac{ m_{-}}{ m \lambda} \left( \frac{ 1 - 2 \chi }{1 - \chi} \right) \right).
									\end{equation*}
								\item If we look for a saddle point $z > \frac{ m_{-}}{A}$, we arrive to:
									\begin{equation*}
									z = \frac{ 2 f - 1 - \chi }{1 - 2 \chi} ,
								\end{equation*}
								which is the solution for
									\begin{equation*}
										f > \frac{1}{2} \left(1  + \frac{ m_{-}}{ m \lambda} \left( \frac{ 1 - 2 \chi }{1 - \chi} \right) \right) + \frac{\chi}{2} .
									\end{equation*}
									Again we notice that there is a region of width $\frac{\chi}{2}$ where no solution is allowed.
							\end{enumerate}
						\item Otherwise, when $\chi > 1 - \frac{m_{-}}{m_{+}}$, phase III appears for $\frac{m_{+}}{m} < \lambda < \frac{m_{-}}{m \left( 1 - \chi \right)}$. The saddle point is univocally determined:
							\begin{equation*}
								z = \frac{ m \lambda}{2 m_{-} - m_{+}} \left( 2 f - 1 \right) .
							\end{equation*}
					\end{enumerate}
				\item In the next phase, $\frac{m_{-}}{A} < 1$ implies the existence of two sub-regions where the saddle point can be placed.
					\begin{enumerate}[(i)]
						\item When $z < \frac{m_{-}}{A}$ only the second $\theta$ function contributes and the saddle point is:
							\begin{equation*}
								z = \frac{ m \lambda}{ 2 m \lambda \left( 1 - \chi \right) - m_{+}} \left( 2 f - 1  \right).
							\end{equation*}
							This result is self-consistent for
							\begin{equation*}
								f < \frac{1}{2} \left[ 1 + 2 \frac{ m_{-}}{m \lambda } - \frac{ m_{+} m_{-} }{m^2 \lambda^2 \left( 1 - \chi \right)} \right] .
							\end{equation*}
						\item When $z > \frac{m_{-}}{A}$ both $\theta$ functions enter the calculation and we are led to:
							\begin{equation*}
								z = \frac{ m \lambda}{ 2 m \lambda \left( 1 - \chi \right) - m_{+}} \left( 2 f - 1 - \chi \right) ,
							\end{equation*}
							associated to the condition:
							\begin{equation*}
								f > \frac{1}{2} \left[ 1 + 2 \frac{ m_{-}}{m \lambda } - \frac{ m_{+} m_{-} }{m^2 \lambda^2 \left( 1 - \chi \right)} \right]  + \frac{\chi}{2} .
							\end{equation*}
					\end{enumerate}
				\item The fifth phase is always characterized by symmetric domain, with both resonances lie in the interior of the domain, thus there are three sub-regions to look for saddle points.
					\begin{enumerate}[(i)]
						\item The first sub-region corresponds to $z<- \frac{m_{+}}{A}$:
							\begin{equation*}
								z= \frac{ 2 f - 1 + \chi }{1 - \chi } .
							\end{equation*}
							This is the solution to be considered for $f < \frac{ \lambda - m_{+}/m}{2 \lambda} - \frac{ \chi}{2}$.
						\item In the sub-region $- \frac{m_{+}}{A} < z < \frac{m_{-}}{A}$ the saddle point is:
							\begin{equation*}
								z = \frac{2 f - 1}{1 - \chi} ,
							\end{equation*}
							arising for $\frac{ \lambda - m_{+}/m}{2 \lambda} < f < \frac{ \lambda - m_{-}/m}{2 \lambda}$.
						\item Eventually, when $z >\frac{m_{-}}{A}$ the saddle point becomes:
							\begin{equation*}
								z= \frac{ 2 f - 1 - \chi }{1 - \chi } .
							\end{equation*}
							This is the solution when $f > \frac{ \lambda - m_{-}/m}{2 \lambda} + \frac{\chi}{2}$.
					\end{enumerate}
					We notice that the result of this last phase is analogous to the last phase of Section \ref{sect:WLmassL}, except for the splitting of the mass $m \mapsto m_{\pm}$.
			\end{enumerate}

		\subsubsection{Evaluation of Wilson loop for $\zeta < m_1$}
		\label{sec:FIWL2fameval}
			At this point we ought to use the results of the previous subsection to evaluate the Wilson loop in each phase. By means of standard approximations in the decompactification limit, we find the general saddle point contribution:
			\begin{equation}
			\label{eq:gensaddlepointexpFI}
				S(z_s) = \left[ \frac{ \left( B +  A z \right)^2}{4 A m \lambda} + \cp^{-} \left( z - \frac{m_{-}}{A} \right) \cdot \theta \left( z - \frac{m_{-}}{A} \right) + \cp^{+} \left( z + \frac{m_{+}}{A} \right) \cdot \theta \left( z + \frac{m_{+}}{A} \right) - f z  \right]_{z = z_s} ,
			\end{equation}
			where we have denoted by $z_s$ the saddle point in each phase. We now pass to the phase-by-phase study.
			\begin{enumerate}[(I)]
				\item In the first phase the evaluation of \eqref{eq:gensaddlepointexpFI} simply gives $f (1-f)$, as in the $\zeta =0$ case of Section \ref{sect:WLmassL}. The Wilson loop is therefore:
					\begin{equation}
						\langle W (f) \rangle ^{(0)} \vert_{\mathrm{I}} = e^{m \lambda R N f \left( 1 - f \right)} .
					\end{equation}
				\item Evaluating expression \eqref{eq:gensaddlepointexpFI} in phase II provides the Wilson loop:
					\begin{equation}
						\langle W (f) \rangle ^{(0)} \vert_{\mathrm{II}} = e^{m \lambda R N  \left( \frac{m_{-}}{ 2 m_{-} - m\lambda } \right)^2 \left[ \left(3 - \frac{ 2 m \lambda}{m_{-}} \right) f \left( 1-f \right) + \frac{1}{4} \left( 1 -  \frac{m \lambda }{m_{-}}  \right)^2 \right] } .
					\end{equation}
				\item Wilson loop in phase III takes two different forms, depending on the value of $\chi$.
					\begin{enumerate}[(a)]
						\item If $\chi < 1 - \frac{m_{-}}{m}$ then phase IIIa is realized, and
							\begin{equation*}
								\langle W (f) \rangle ^{(0)} \vert_{\mathrm{IIIa}} = e^{ m \lambda RN \frac{(1- \chi)}{\left( 1 - 2 \chi \right)^2} \left[ \left(1 - 3 \chi \right) f \left( 1 - f \right) - \frac{ 1 - 2 \chi}{4 \left( 1 - \chi \right)} \right] }
							\end{equation*}
							for $f< \frac{1}{2} \left( 1+ \frac{ m_{-}}{ m \lambda} \left( \frac{ 1 - 2 \chi }{1 - \chi} \right) \right)$, or instead giving:
							\begin{equation*}
								\langle W (f) \rangle ^{(0)} \vert_{\mathrm{IIIa}} = e^{ m \lambda RN \frac{(1-\chi) }{\left( 1 - 2 \chi \right)^2} \left[  \left(1 - 3 \chi^2 \right) f \left( 1 - f \right) + \frac{1}{2} \left( 1 - 2 \chi \right)\left( 1 + \chi \right)^2 + \frac{\left( 1 - 2 \chi \right)}{4 \left( 1 - \chi \right)}\left( 1 - 2 \chi \frac{m_{-}}{m \lambda} \right) \right] }
							\end{equation*}
							for $f> \frac{1}{2} \left( 1+ \frac{ m_{-}}{ m \lambda} \left( \frac{ 1 - 2 \chi }{1 - \chi} \right) \right) + \frac{\chi}{2}$.
						\item For values of $\chi > 1 - \frac{m_{-}}{m}$ phase IIIb arises, leading to the Wilson loop
							\begin{equation}
								\langle W (f) \rangle ^{(0)} \vert_{\mathrm{IIIb}} = e^{ m \lambda R N \left( \frac{ m_{-}}{2 m_{-} - m_{+}} \right)^2  \left[ \left(3 - \frac{2 m_{+}}{m_{-}} \right) f \left( 1 - f \right) - \frac{1}{4} \left( 3 - \frac{2 m_{+}}{m_{-}} \right) + \frac{m_{+}}{m \lambda}  \left( 1 - \frac{m_{+}}{2 m_{-}} \right)^2 \left( 2 - \frac{m_{+}}{m \lambda} \right) \right] } .
							\end{equation}
							One can easily prove that in the $\zeta \to 0$ limit the expression converges to the Wilson loop in middle phase from Section \ref{sect:WLmassL}.
					\end{enumerate}
				\item Depending on the relation between $f$ and the parameters $\lambda, \chi$ we have two possible saddle point solutions in phase IV, one or the other determining the Wilson loop, which can be expressed as:
							\begin{equation}
								\langle W (f) \rangle ^{(0)} \vert_{\mathrm{IV}} = e^{ m \lambda R N \left( \frac{ m \lambda \left( 1 - \chi \right) }{2 m \lambda \left(1-\chi \right) - m_{+}} \right)^2  \left[ \left(3 - \frac{2 m_{+}}{m \lambda \left( 1 - \chi \right)} \right) f \left( 1 - f \right) + \psi_{\lambda} (f) \right] } ,
							\end{equation}
							where we have introduced the auxiliary function $\psi_{\lambda}$:
							{\small \begin{equation*}
								\psi_{\lambda} (f) =   - \frac{1 + \epsilon(f) \chi }{4} \left( 3 - \frac{2 m_{+}}{m \lambda \left( 1 - \chi \right)} \right) + \frac{m_{+}}{m \lambda}  \left( 2 - \frac{m_{+}}{m \lambda}  \right) \left( 1 - \frac{1}{2} \left(\frac{m_{+}}{m \lambda \left( 1 - \chi \right)} \right)\right) ^2 ,
							\end{equation*}}
							$\epsilon (f) = 0 $ for $f < \frac{1}{2} \left[ 1 + 2 \frac{ m_{-}}{m \lambda } - \frac{ m_{+} m_{-} }{m^2 \lambda^2 \left( 1 - \chi \right)} \right]$ and $1$ for $ f > \frac{1}{2} \left[ 1 + 2 \frac{ m_{-}}{m \lambda } - \frac{ m_{+} m_{-} }{m^2 \lambda^2 \left( 1 - \chi \right)} \right] + \frac{\chi}{2}$.
				\item Eventually in phase V there are three sub-regions, associated to three (disconnected) intervals for $f$. Evaluating the Wilson loop for saddle point in each sub-region we arrive to:
					\begin{equation}
						\langle W (f) \rangle ^{(0)} \vert_{\mathrm{V}} = e^{ m \lambda R N  \left[ f \left( 1  - f \right) - \chi \widetilde{\beta} _{\lambda} (f) \right] } ,
					\end{equation}
					with auxiliary function $\widetilde{\beta} _{\lambda} (f)$ reducing to the $\beta_{\lambda} (f)$ defined in \eqref{eq:betashortcut} as $\zeta \to 0$. Explicitly:
				\begin{equation}
						\widetilde{\beta} _{\lambda} (f) = \begin{cases} f , \\ \frac{\lambda - m_{+}/m }{2 \lambda} - \frac{\chi}{4} ,  \\  1 - f + \frac{ m_{+} - m_{-}}{2 m \lambda} ,  \end{cases} \ \begin{matrix} \ & f < \frac{\lambda - m_{+}/m}{2 \lambda} , \\  \frac{\lambda - m_{+}/m}{2 \lambda} < & f <\frac{\lambda + m_{-}/m}{2 \lambda} ,  \\ \  & f > \frac{\lambda + m_{-}/m}{2 \lambda} . \end{matrix}
					\end{equation}
					The Wilson loop in the final phase is therefore analogous to the the result of Section \ref{sect:WLmassL}, but the FI term modifies the last summand consistently with the splitting of the mass $m \mapsto m_{\pm}$.
			\end{enumerate}\par
			\medskip
			The order of the phase transition is given by the derivatives of the corresponding free energy, which we recall is defined as:
			\begin{equation*}
				\mathcal{W} = - \frac{1}{RN} \log \langle W \left( f \right) \rangle .
			\end{equation*}
			First order derivatives of $\mathcal{W}$ result in cumbersome expressions, given in equation \eqref{eq:WL1d2ms} of Appendix \ref{app:WLfirstderiv}. The derivatives are continuous at critical values between phase I and II and between phase IV and V, this latter by taking $f$ around $\frac{1}{2}$, that is solution (ii) must hold for both phases. Nevertheless the expressions in middle phases fail to be continuous at critical values.\par
			Therefore the systems presents two transitions of second order and two of first order. More specifically: from phase I to II is of second order, from phase II to III and from III to IV is of first order, and finally from phase IV to V is of second order. This difference with the case without FI parameter arises when the eigenvalue density has non-symmetric domain $B \ne A$ at critical values. This is due to the fact that the exponent appearing in the Wilson loop yields a prefactor $\left( \frac{A}{2A-B} \right)^2$. Then, the effect of the asymmetry in the domain is that a non-trivial dependence on $\lambda$ remains. This fact produces a mismatch since the first derivative. Conversely, if the critical value is such that the domain is symmetrical, $B=A$, the prefactor simplifies and the transition is of second order.

		\subsection{Antisymmetric Wilson loop with one mass scale}
		\label{sec:FIWL1fameval}
				
			To complete the discussion, we now obtain the Wilson loop in antisymmetric representation in the decompactification limit, in the case of only one set of hypermultiplets, with unconstrained mass $-m_1$. That is, we set $\nu_{-} =1 , \nu_{+}=0$ and refer to results of Subsection \ref{sect:zsmall1fam}, but, as we have seen, all other cases such as $\zeta >m_1$ or exchanging values of $\nu_{+}$ and $\nu_{-}$ can be easily recovered.\par
			The saddle point equation in this case carries only one theta function:
			\begin{equation}
			\label{eq:gensaddlepoint1fam}
				z= \frac{ 2 m \lambda}{2 A - B } \left( f -  \cp ^{-} \theta  \left( z - \frac{m_{-}}{A} \right)  \right) - \frac{B}{2A - B} ,
			\end{equation}
			and the saddle point is obtained in each phase from this latter expression.
			\begin{enumerate}[(I)]
				\item \begin{equation*}
						z= \frac{ 2 f - 1 + \frac{\chi}{2} }{1 + \frac{3\chi}{2} } .
					\end{equation*}
				\item \begin{equation*}
						z = \frac{ m \lambda }{ 2 m_{-} - m \lambda \left( 1 - \frac{\chi}{2} \right) } \left( 2 f - 1 + \frac{\chi}{2} \right) .
					\end{equation*}
				\item In phase III the saddle point can be placed in two sub-regions, namely on the left or on the right of $\frac{m_{-}}{m \lambda \left( 1 - \frac{\chi}{2} \right)}$. Therefore one arrives to:
					\begin{enumerate}[(i)]
						\item \begin{equation*} 
							z= \frac{2 f - 1 + \frac{\chi}{2} }{1 - \frac{\chi}{2}} 
						\end{equation*}
						for $f < \frac{1}{2}\left( \frac{m_{-}}{m \lambda} + 1 \right) - \frac{\chi}{4} $ ,
						\item or instead
						\begin{equation*}
							z = \frac{2 f - 1 - \frac{\chi}{2} }{1 - \frac{\chi}{2}} 
						\end{equation*}
						for $f>\frac{1}{2}\left( \frac{m_{-}}{m \lambda} + 1 \right) + \frac{\chi}{4} $.
					\end{enumerate}
					We remark that again for values of $f$ in an interval of width $\frac{\chi}{2}$ no consistent saddle point is found.
			\end{enumerate}
			Standard procedure, identical to the previous case, leads to the results:\begingroup
\renewcommand*{\arraystretch}{1.2}
			\begin{equation}
			\begin{aligned}
				& \langle W (f) \rangle ^{(0)} \vert_{\mathrm{I}} =  e^{ m \lambda R N  \frac{ \left( 1 + \frac{5 \chi}{2} \right) \left( 1 + \frac{\chi}{2} \right) }{ \left( 1 + \frac{3 \chi}{2} \right)^2 } \left[ f \left( 1 - \frac{\chi}{2} - f \right)  + \frac{\chi^2 \left( 1 - \frac{\chi}{2} \right)^2}{ 4 \left( 1 + \frac{5 \chi}{2} \right) \left( 1 + \frac{\chi}{2} \right) } \right] }  , \\
				& \langle W (f) \rangle ^{(0)} \vert_{\mathrm{II}} =  e^{ m \lambda R N \left( \frac{ m_{-} }{2 m_{-} - m \lambda \left( 1 - \frac{\chi}{2} \right) } \right)^2 \left[ f \left( 1 - \frac{\chi}{2} - f \right) \left( 3 - 2 \frac{ m \lambda \left(1 - \frac{\chi}{2} \right)}{m_{-}} \right) + \left( 1 - \frac{\chi}{2} \right)^2 \left( 1 -  \frac{ m \lambda \left(1 - \frac{\chi}{2} \right)}{ m_{-}} \right)^2 \right] } , \\
				& \langle W (f) \rangle ^{(0)} \vert_{\mathrm{III}} = \begin{cases}  e^{m \lambda R N  f \left( 1 - \frac{\chi}{2} - f \right) } , \\ e^{ m \lambda R N  \left[ f \left( 1 + \frac{\chi}{2} - f \right)   - \frac{\chi }{2} \left( 1 + \frac{ m_{-}}{m \lambda} \right) \right] } , \end{cases} \ \begin{matrix} f < & \frac{1}{2}\left( \frac{m_{-}}{m \lambda} + 1 \right) - \frac{\chi}{4} , \\ f > & \frac{1}{2}\left( \frac{m_{-}}{m \lambda} + 1 \right) + \frac{\chi}{4} . \end{matrix}
			\end{aligned}
			\end{equation}\endgroup
			Wilson loop is continuous assuming $f$ such that solution (i) holds in phase III. We take derivatives of the logarithm of those expressions, which are given in \eqref{eq:WL1d1ms} in Appendix \ref{app:WLfirstderiv}, obtaining that the phase transition from I to II is first order, whilst the transition from phase II to III is second order. As for the case with two mass scales, the phase transition is first order at that critical value for which the domain is not symmetric, $A \ne B$, while at critical value for which $A=B$ simplifications occur, and the transition is second order.

	\section{Outlook}
		 
	It would be interesting to also study the large $N$ behavior of Wilson loops in the symmetric representation, as has been done for four dimensional theories. In this concluding section, we briefly mention about this case along the lines of the work carried out here. \par
	We know from \cite{Hartnoll:2006} that the Wilson loop in the symmetric representation of rank $k$ is given by the expression:
	\begin{equation}
			\langle W_k \rangle = \oint_{\alpha} \frac{ \dd w }{2 \pi \I } \frac{1}{ w ^{1+k}} \exp \left\{ - N \int_{-A} ^B \dd x \rho (x) \log \left( 1 - w e^{x} \right) \right\} ,
	\end{equation}
	where, as in Section \ref{sect:WLmassL}, $\alpha$ is a circle around the origin. After the same change to cylindric variables we did at the beginning of Section \ref{sect:WLmassL}, and using $f=k/N$, we arrive to:
	\begin{equation}
			\langle W (f) \rangle = \frac{ AR }{2 \pi \I } \oint_{\gamma} \dd z \exp \left\{ - A R N \left[ \int_{-B/A} ^{1} \frac{ \dd y }{R} \rho \left( - A y \right) \log  \left( 1 - e^{- AR \left( y - z \right) } \right) + f z \right] \right\} .
	\end{equation}
	Although this formula is similar to \eqref{eq:WLgeneralformbeforeSP} for the antisymmetric case, the different sign in the logarithm is crucial. Indeed, as was noticed in \cite{Hartnoll:2006}, for the symmetric case the branch cut of the logarithm lies on $\im z = 0$. As a consequence, $\gamma$ wraps the cylinder at $\re z < - \frac{B}{A}$. This is because $\gamma$ comes from $ - \infty$ as we increase the radius of the original contour $\alpha$, but we cannot go further $- \frac{B}{A}$ due to the discontinuity of the logarithm. If we want to pursue saddle point in the symmetric case, we have to take into account the jump from a Riemann sheet to another. However, if we try to do so in the same manner as above we see that the procedure does not apply straightforwardly, and hence we shall discuss this other problem elsewhere.\par
	\medskip
	As we have seen throughout this work, the presence of the FI parameter increases the number of phase transitions, due to the ``splitting'' of the masses. In this way, we have seen that $S$ mass scales lead to $2S+1$ different phases. In addition, it introduces the possibility of having asymmetry in the interval, which is the compact support of the eigenvalue density of the matrix model, for large $N$. As we have seen at the end, this is responsible for having phase transitions of the Wilson loops at first order instead of at second order. This is a general mechanism that seemingly will extend to the antisymmetric Wilson loops studied in \cite{Russo:2017ngf}. It would be interesting to explicitly check this, by adding a FI parameter in \cite{Russo:2017ngf} and doing the corresponding analysis.
	Finally, it is worth mentioning that at first apparently unrelated works in statistical mechanics, study in fact similar problems and systems \cite{Cunden:2017oja,Dhar:2017grt}.

	\acknowledgments
		We specially thank Luis Melgar, who shared with us long ago a set of notes on this problem. Thanks also to Jorge Russo for valuable comments on the draft. The work of MT was supported by the Fundação para a Ciência e a Tecnologia through its program Investigador FCT IF2014, under contract IF/01767/2014. The work of LS was supported by the Fundação para a Ciência e a Tecnologia through the doctoral scholarship SFRH/BD/129405/2017.

	\begin{appendix}
	
	\section{Eigenvalue densities and free energies}
		We collect here some calculations that complete the discussion of Section \ref{sect:FItermdistrib}.
	
		\subsection{Further eigenvalue densities}
		\label{app:tablerho}
			Below we schematically present other eigenvalue densities. The three cases correspond, respectively, to a single set of hypermultiplets with mass $-m_1>\zeta$, two equal sets of hypermultiplets with masses $m_2$ and $-m_1< \zeta$, and two different sets of hypermultiplets $\nu_{-} \ne \nu_{+}$ with masses $m_2, -m_1>\zeta$. In the last case we also defined $\nu_0 = \frac{ \nu_{-} - \nu_{+}}{2}$ and $\nu = \frac{\nu_{-} + \nu_{+}}{2}$. The coefficients $\cp ^{\pm}, A,B$ are defined in \eqref{eq:rhogenP}.
			\begin{figure}[hbt]
			\centering
			\includegraphics[width=0.7\textwidth]{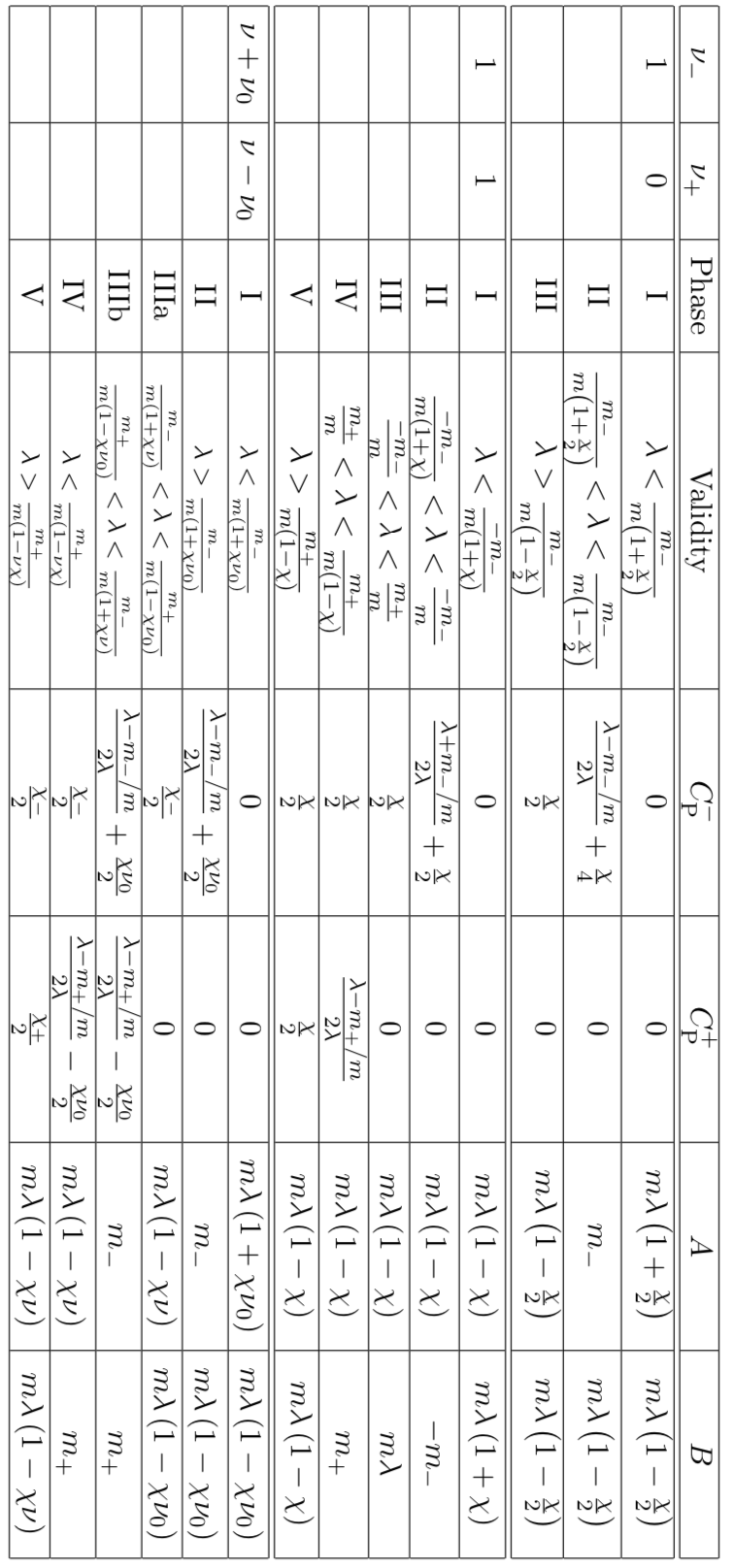}
			\caption{Details of eigenvalue densities in various cases.}
			\end{figure}
			\clearpage

		\subsection{Free energy derivatives}
		\label{app:freeen2d}
			In this appendix we provide explicit expressions for derivatives of the free energy in the cases described throughout the text.
			
			\subsubsection{$\zeta < m_1$ with two mass scales}
				When two mass scales of opposite sign are present, with small FI coupling $\zeta < m_1$, first order derivative of the free energy is given in expression \eqref{eq:firstorderFEa2}, when $\chi > 1 - \frac{m_{-}}{m}$, and in expression \eqref{eq:firstorderFEa1} when $\chi < 1 - \frac{m_{-}}{m}$ . From that we get:\begingroup
\renewcommand*{\arraystretch}{1.2}
				\begin{equation*}
					\frac{ \partial^2 \fee }{ \partial \lambda^2}   = \begin{cases} 0 ,  \\ \frac{m}{2} \left[ \lambda^{-3} \left( \frac{m_{-} }{m} \right)^2 - \lambda^{-4} \left( \frac{m_{-} }{m} \right)^3 \right] , \\ \frac{m}{2} \left[ \lambda^{-3} \left(  \frac{ m_{-}^2 + m_{+} ^2}{m^2} \right) -  \lambda^{-4} \left( \frac{ m_{-}^3 + m_{+} ^3 }{m^3} \right) \right] , \\  \frac{m}{2} \left[ \lambda^{-3} \left( \frac{\chi  m_{-}^2 + m_{+} ^2 }{m^2 }   \right) - \lambda^{-4} \left( \frac{m_{+} }{m} \right)^3 \right] , \\  \frac{m}{2} \chi \lambda^{-3} \left( \frac{  m_{-}^2 + m_{+} ^2 }{m^2} \right) , \end{cases}   \ \begin{matrix} \ & \quad \lambda < & \frac{ m_{-} }{m} , \\ \frac{ m_{-} }{m} & < \lambda < & \frac{m_{-}}{m \left( 1 -  \chi \right) } , \\  \frac{m_{-}}{m \left( 1 -  \chi \right)} & <  \lambda < & \frac{ m_{+}}{m} , \\ \frac{m_{+}}{m} & <  \lambda < & \frac{m_{+}}{m \left( 1 - \chi \right)} , \\ \ & \quad \lambda > & \frac{ m_{+} }{ m \left( 1 - \chi \right)} .  \end{matrix}
				\end{equation*}\endgroup
				for $\chi > 1 - \frac{m_{-}}{m_{+}} $, while for $\chi < 1 - \frac{m_{-}}{m_{+}} $ one simply has to replace the value of phase III and exchange the role of the critical points:
				\begin{equation*}
				     \frac{ \partial^2 \fee }{ \partial \lambda^2} \rvert_{\mathrm{III}} = \frac{m}{2} \chi \lambda^{-3} \left( \frac{m_{-}}{m} \right)^2 , \qquad  \frac{ m_{+}}{m} <  \lambda < \frac{m_{-}}{m \left( 1 -  \chi \right)} .
				\end{equation*}
				The latter expressions are continuous but their derivatives are not, hence phase transitions are of third order.
				
			\subsubsection{$\zeta < m_1$ with one mass scale}
				In the case of only one mass scale, first order derivative of the free energy is:
				\begingroup
\renewcommand*{\arraystretch}{1.2}
				\begin{equation*}
					 \frac{ \partial \fee }{ \partial \lambda} = \begin{cases} - \frac{m}{6} \left[ 1 + 3 \left( \frac{\chi}{2} \right)^2 \right],  \\ - \frac{m}{12} \left[ \left( 1 - \frac{ \chi }{2} \right)^3 +3 \left( 1 + \frac{ \chi }{2} \right) \lambda^{-2} \left( \frac{m_{-} }{m} \right)^2 - 2 \lambda^{-3} \left( \frac{m_{-} }{m} \right)^3 \right] ,  \\ - \frac{m}{6} \left[ \left( 1 - \frac{ \chi }{2} \right)^3 +3 \left( \frac{ \chi }{2} \right) \lambda^{-2} \left( \frac{m_{-} }{m} \right)^2  \right] , \end{cases} \ \begin{matrix} \ & \quad \lambda < & \frac{m_{-} }{m \left( 1 + \frac{\chi }{2} \right) } , \\ \frac{m_{-} }{m \left( 1 + \frac{\chi }{2} \right) } & < \lambda < & \frac{m_{-} }{m \left( 1 -\frac{\chi }{2} \right) } , \\   \ & \quad \lambda > & \frac{m_{-} }{m \left( 1 -\frac{\chi }{2} \right) } .  \end{matrix}
				\end{equation*}\endgroup
				Second order derivative is:\begingroup
\renewcommand*{\arraystretch}{1}
				\begin{equation*}
					\frac{ \partial^2 \fee }{ \partial \lambda^2} = \begin{cases} 0 ,  \\ \frac{m}{2} \left[ \left( 1 + \frac{\chi}{2} \right) \lambda^{-3} \left( \frac{ m_{-} }{m} \right)^2 - \lambda^{-4} \left( \frac{ m_{-} }{m} \right)^3  \right] , \\ m \left( \frac{ \chi }{2} \right) \lambda^{-3} \left( \frac{ m_{-} }{m} \right)^2 , \end{cases}  \ \begin{matrix} \ & \quad \lambda < & \frac{m_{-} }{m \left( 1 + \frac{\chi }{2} \right) } , \\ \frac{m_{-} }{m \left( 1 + \frac{\chi }{2} \right) } & < \lambda < & \frac{m_{-} }{m \left( 1 -\frac{\chi }{2} \right) } , \\   \ & \quad \lambda > & \frac{m_{-} }{m \left( 1 -\frac{\chi }{2} \right) } ,  \end{matrix}
				\end{equation*}\endgroup
				which again is a continuous function of $\lambda$, but with discontinuous derivative.
				
			\subsubsection{$\zeta > m_1$ with two mass scales}
				When the system includes two mass scales with both resonances at positive eigenvalues, the first order derivative of the free energy is:
				\begingroup
\renewcommand*{\arraystretch}{1.35}
				\begin{equation*}
					\frac{ \partial \fee }{ \partial \lambda} = \begin{cases} - \frac{m}{6} \Big[ 1 + 3 \chi^2 \Big] ,  \\ - \frac{m}{12} \left[ \left( 1 - \chi \right)^3 + 3  \left( 1 + \chi \right) \lambda^{-2} \frac{ m_{-} ^2}{m^2}   - 2 \lambda^{-3}  \frac{ -m_{-} ^3}{m^3}  \right] , \\ - \frac{m}{12} \left[ 1+ \left( 1 - \chi \right)^3 + 3 \chi \lambda^{-2}  \frac{m_{-}^2}{m^2}  \right] , \\  - \frac{m}{12} \left[ \left( 1 - \chi \right)^3 + 3 \lambda^{-2} \left(  \frac{ \chi  m_{-} ^2 + m_{+} ^2 }{m^2}   \right) - 2 \lambda^{-3}   \frac{ m_{+} ^3}{m^3}  \right] ,   \\  - \frac{m}{12} \left[ 2 \left( 1 - \chi \right)^3 + 3 \chi \lambda^{-2} \left(  \frac{ m_{-} ^2 + m_{+} ^2}{m^2}   \right) \right] ,  \end{cases} \ \begin{matrix} \ & \quad \lambda < & \frac{ - m_{-} }{m \left( 1 + \chi \right) } , \\ \frac{ - m_{-} }{m \left( 1 + \chi \right) } & < \lambda < &  \frac{ - m_{-}}{m} ,  \\ \frac{ - m_{-}}{m} & < \lambda < & \frac{ m_{+}}{m} , \\ \frac{m_{+}}{m} & < \lambda < & \frac{m_{+}}{m \left( 1 - \chi \right)} , \\ \ & \quad \lambda > & \frac{m_{+}}{m \left( 1 - \chi \right)} . \end{matrix}
				\end{equation*}\endgroup
				Second order derivative is then:\begingroup
\renewcommand*{\arraystretch}{1.3}
				\begin{equation*}
					\frac{ \partial^2 \fee }{ \partial \lambda^2} = \begin{cases} 0 \\ \frac{m}{2} \left[ \left( 1 + \chi \right) \lambda^{-3}  \frac{ m_{-}^2 }{m^2}   - \lambda^{-4}  \frac{ - m_{-}^3}{m^3}   \right]  \\  \frac{m}{2} \left[ \chi \lambda^{-3}  \frac{ m_{-}^2 }{m^2}  \right] \\ \frac{m}{2} \left[ \lambda^{-3}  \left( \frac{ \chi m_{-}^2  m_{+} ^2 }{m^2}  \right)- \lambda^{-4}   \frac{ m_{+}^3}{m^3}  \right] \\ \frac{m}{2} \chi \lambda^{-3} \left[  \frac{ m_{-}^2 + m_{+}^2 }{m^2}   \right] ,  \end{cases}  \ \begin{matrix} \ & \quad \lambda < & \frac{ - m_{-} }{m \left( 1 + \chi \right) } , \\ \frac{ - m_{-} }{m \left( 1 + \chi \right) } & < \lambda < &  \frac{ - m_{-}}{m} ,  \\ \frac{ - m_{-}}{m} & < \lambda < & \frac{ m_{+}}{m} , \\ \frac{m_{+}}{m} & < \lambda < & \frac{m_{+}}{m \left( 1 - \chi \right)} , \\ \ & \quad \lambda > & \frac{m_{+}}{m \left( 1 - \chi \right)} , \end{matrix}
				\end{equation*}\endgroup
				which is continuous at all critical values of $\lambda$. Moreover, third order derivative is discontinuous at any of those critical points, meaning that phase transitions are all third order.

	\section{Matrix model: technical aspects}
	
		In this appendix we comment on some technical aspects of the matrix model of interest. In \ref{app:PTmechanism} we analyze the type of phase transition that occurs and present the differences with other known third order phase transitions in the literature. In appendix \ref{app:holomorphicMM} we present the framework for the study of the case of complex masses at finite radius.
	
		\subsection{Comments on the phase transition mechanism}
		\label{app:PTmechanism}
		Third order phase transitions are a widespread phenomenon in gauge theories that admit a matrix model description, the most celebrated being the Gross--Witten--Wadia (GWW) transition in lattice two-dimensional Yang--Mills theory \cite{Gross:1980he,Wadia:1980cp}, and the Douglas--Kazakov (DK) one in two-dimensional Yang--Mills theory on the sphere \cite{Douglas:1993iia}. However, it is worth stressing that not all of them are triggered by the same mechanism, and different classes of theories showing third order phase transition can be identified depending on the underlying mechanism.\par
		For example, both the GWW and the DK transition arise due to some additional constraint imposed on the eigenvalue density $\rho$. The nature of those constraints is nevertheless different: the GWW model is a unitary matrix model, and the compactness imposes a maximum distance between two eigenvalues, leading to the condition $\rho \ge 0$. Conversely, the DK matrix model is a discrete Hermitian ensemble, and the discreteness imposes a minimum distance between eigenvalues, leading to the condition $\rho \le 1$. At values of the 't Hooft coupling for which the eigenvalue density reaches the boundary of the inequality, the system undergoes a phase transition. A phase transition can also be caused by the presence of hard walls in the Coulomb gas description: the eigenvalues are free as long as the support of the eigenvalue density is far from the wall, but when the support hits the wall the system undergoes a phase transition, which, under mild assumptions, is third order \cite{Cunden:2017oja}.\par
		In all the mentioned cases, the phase transition admits a characterization in terms of the support of the eigenvalue density. In the GWW model, the transition corresponds to a gap opening in the support, thus below the critical point the eigenvalues distribute all over the circle while above the critical point they distribute only along an arc. In the DK case, the solution passes from one-cut to two-cut, with a saturated region $\rho=1$ in the middle. In presence of hard walls, the transition corresponds to soft-edge to hard-edge transition, and the latter may also include a saturated region close to the wall, if the matrix model is discrete.\par
		The present work, however, is inserted in a different stream of research \cite{Russo:2013qaa,Russo:2013kea,Nedelin:2015mta}, in which the decompactification limit plays a crucial role. In those cases, the transition does not arise from a constraint on the eigenvalue density coming from the specifics of the random matrix ensemble. Instead, it appears at the level of saddle point equation. This is a consequence of the fact that, in the large radius limit, the derivative of the potential, namely $V^{\prime}$, becomes discontinuous. The eigenvalue density is determined, through the saddle point equation, in terms of $V^{\prime}$, and its support is, in general, a function of the scaling parameters. For values of those parameters such that the boundaries of the support hit a discontinuity of $V^{\prime}$, the solution to the saddle point equation changes and the system undergoes a phase transition.

		\subsection{Comments on holomorphic matrix models}
		\label{app:holomorphicMM}
		
			As explained in the main text, Section \ref{sec:complexmasses}, to study the large $N$ limit of the model with complex masses at finite radius, we have to pass from an Hermitian matrix model, with eigenvalues on the real line, to an holomorphic matrix model, with eigenvalues distributed along a curve $\Gamma \subset \mathbb{C}$ \cite{Lazaroiu:2003vh}.\par
\begin{figure}[hbt]
						\centering
						\includegraphics[width=0.5\textwidth]{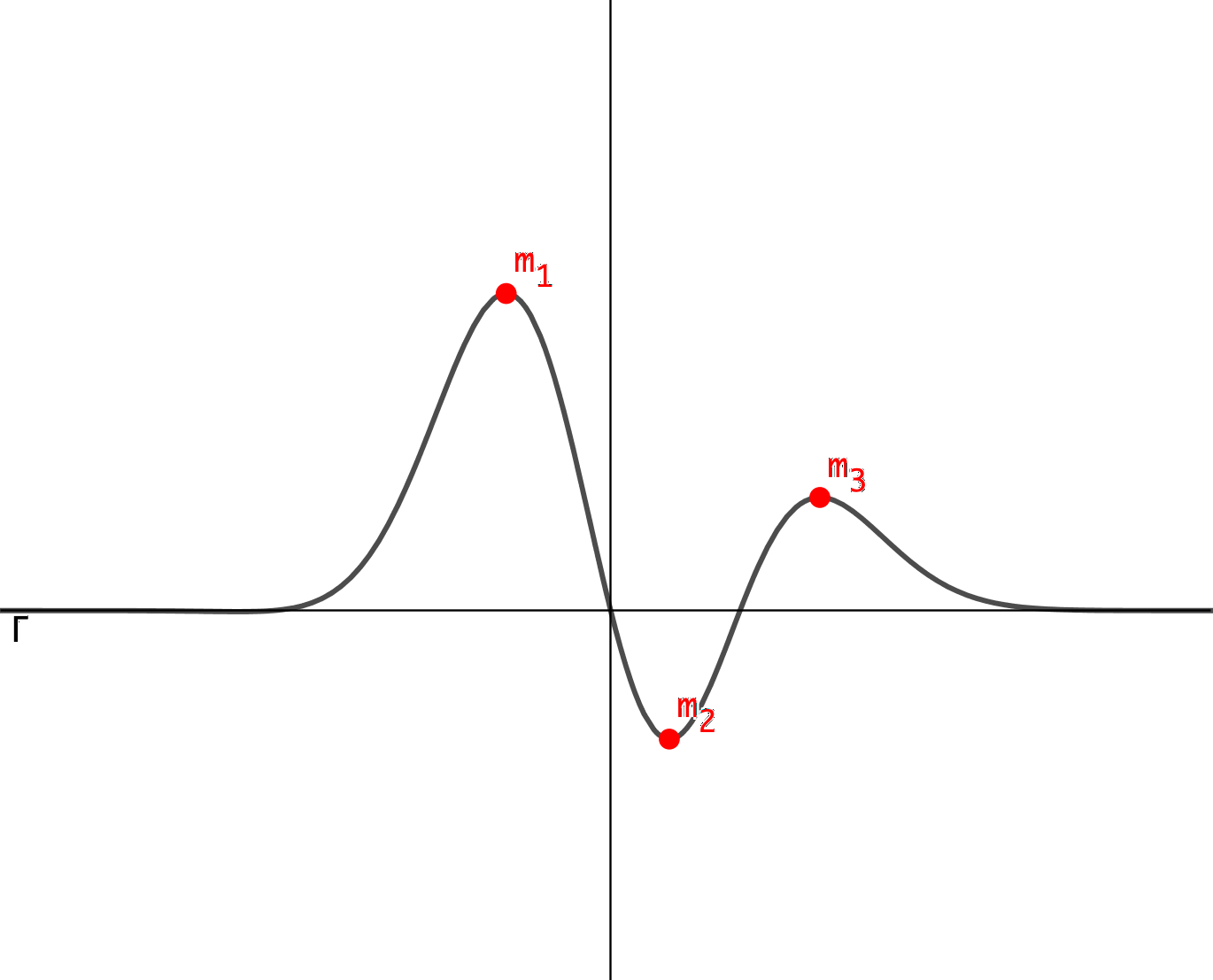}
						\caption{Integration contour in the complex plane, for the holomorphic matrix model.}
					\label{fig:Gammacurve}
				\end{figure}

			The first issue we have to address is how to choose the curve $\Gamma$, as the resulting matrix model will of course depend on such choice, up to homotopy. Indeed, holomorphicity of the integrand guarantees that the result only depends on the homotopy class of $\Gamma$, and all paths $\Gamma$ that avoid the poles at $m_{\mathbb{C},a} \pm i \pi (\mathrm{mod} \ i 2 \pi )$ in a prescribed way provide the same result. A suitable choice, which preserves the convergence of the integral and reproduces the correct answer as the masses approach the real line, is a curve interpolating between the values of the masses in the complex plane, without self-intersections and asymptotically approaching the real axis, as sketched in Figure \ref{fig:Gammacurve}.\par
			For the holomorphic version of the matrix model, the saddle point equations can be rearranged into the integral equation:
			\begin{equation}
			\label{eq:complexSPE}
				\dashint \dd s^{\prime} \rho (s^{\prime}) \coth \left( \frac{ x(s)  - x(s^{\prime}) }{2} R \right) = \frac{1}{2 t } V^{\prime} (x (s)) ,
			\end{equation}
			where $s$ is a coordinate along $\Gamma$, $t$ is the 't Hooft coupling and $V^{\prime}$ is the derivative of the potential in the Veneziano limit. Here, $x(s) \in \mathbb{C}$ is a complex eigenvalue placed along $\Gamma$, and the eigenvalue density $\rho (s)$ is a complex function supported on some arcs along $\Gamma$. Requiring $\rho (s)$ to be real-valued would impose further constraints on the choice of $\Gamma$. Complex saddle point equations as \eqref{eq:complexSPE} can be studied following \cite{Lazaroiu:2003vh}. The present analysis and the holomorphic dependence of the partition function on the (complex) masses suggest that, for small imaginary part of the masses, such that the curve $\Gamma$ is homotopic to the real line, the solution could be obtained as a prolongation to complex values of the solution for real masses.

		\section{Regions with no saddle point}
		\label{app:noregSPWL}
			
			In Section \ref{sect:WLmassL} we studied the Wilson loop at leading order in the large radius limit, considering the eigenvalue density with only mass deformation. From general arguments we expected not to find consistent solutions to the saddle point equation \eqref{eq:saddlepoint3integrals} in the region $| \re z | \ge 1$. Here we formally confirm this claim.
			
			\subsubsection{$\re z > 1$ or $\re z < -1$}
				When $\re z > 1$ one has:
				\begin{equation*}
				\begin{aligned}
					& 1 + e^{- AR \left( z \pm 1 \right) } \  \ \xrightarrow{R \to \infty} 1 , \qquad \mathcal{I}_1 \to 2 AR , \\
					& 1 + e^{- AR \left( z + \frac{ m }{A} \right) }  \xrightarrow{R \to \infty} 1 , \qquad \mathcal{I}_2 \to 1 , \\
					& 1 + e^{- AR \left( z - \frac{ m }{A} \right) }  \xrightarrow{R \to \infty} 1 , \qquad \mathcal{I}_3 \to 1 ,
				\end{aligned}
				\end{equation*}
				where the evaluation of the last two integrals only holds in phases II and III where $A \ge m$, and they vanish in phase I. We see that, taking the decompactification limit in this region, the dependence on $z$ is lost, and the saddle point equation reduces to
				\begin{equation*}
					\frac{ A }{m \lambda} + 2 \cp = f ,
				\end{equation*}
				Plugging the expressions for $\cp$ in each phase one sees that the saddle point is only satisfied for $f \equiv 1$.\par
				An identical procedure works for $\re z < -1$, in which case:
				\begin{equation*}
				\begin{aligned}
					\log \left( \frac{ 1 + e^{- AR \left( z + 1 \right) }}{1 + e^{-AR \left( z-1 \right)} } \right) \xrightarrow{R \to \infty}  - 2 AR , \qquad   & \mathcal{I}_1 \to  0 , \\
					1 + e^{- AR \left( z + \frac{ m }{A} \right) } \xrightarrow{R \to \infty} 0 , \qquad & \mathcal{I}_2 \to 0 , \\
					1 + e^{- AR \left( z - \frac{ m }{A} \right) } \xrightarrow{R \to \infty} 0 , \qquad & \mathcal{I}_3 \to 0 ,
				\end{aligned}
				\end{equation*}
				and the saddle point equation reduces to the trivial expression $ f \equiv 0$.\par
				We thus find out that the action is flat in the region $\re z < -1$ and $\re z > 1 $ only if $f$ is exactly $0$ or $1$ respectively, in which case the solution satisfies the reflection symmetry $z \mapsto -z $ with $f \mapsto 1 - f$. 
			
			\subsubsection{$\re z $ around $ \pm 1$}
				We consider the case $\re z = \pm 1 + \oR$ separately. Writing $z = 1 + \frac{z_1}{mR}$, we have:
				\begin{equation*}
					\log \left( \frac{ 1 + e^{- AR \left( z + 1 \right) }}{1 + e^{-AR \left( z-1 \right)} } \right) \xrightarrow{R \to \infty} - \log \left( 1 + e^{\frac{A}{m} z_1 } \right) , \qquad \mathcal{I}_1 = 2 AR - \log \left( 1 + e^{\frac{A}{m} z_1 } \right)  ,
				\end{equation*}
				while $\mathcal{I}_2$ still goes to $1$ in both phase II and III, and $\mathcal{I}_3$ also receives corrections in phase II, where $A=m$, given by
				\begin{equation*}
					\mathcal{I}_3 \vert_{\mathrm{II}} \to \frac{1}{1 + e^{-z_1 } } .
				\end{equation*}
				Therefore, in phase I and III the saddle point is again satisfied only for $f=1$ at leading order in the decompactification limit, but in phase II we are left with
				\begin{equation*}
					1 + \frac{ \lambda - 1}{2 \lambda} \left( 1 + \frac{1}{1 + e^{- z_1 } } \right) = f.
				\end{equation*}
				We can solve this for $z_1$, obtaining:
				\begin{equation}
					z_1 = \log \left( \frac{ \lambda \left( 2 f - 3 \right) + 1 }{2 \lambda \left( 2 - f \right) -2 }\right) .
				\end{equation}
				Recalling that in phase II $\lambda >1$, the solution to the saddle point equation at this order is displaced from the real axis, and cannot be reached because of the branch cut.\par
				We can work out the case $z = - 1 + \frac{z_1}{mR}$ in the same way, exchanging the roles of $\mathcal{I}_2$ and $\mathcal{I}_3$. Then we have
				\begin{equation*}
					\mathcal{I}_1 \to 0 , \quad \mathcal{I}_3 \to 0 ,
				\end{equation*}
				and the contribution of $\mathcal{I}_2$ vanishing in phases I and III but receiving correction in phase II, so that the saddle point equation becomes:
				\begin{equation*}
					\frac{\lambda - 1 }{2 \lambda} \left( \frac{1}{1 + e^{- z_1}} \right) = f .
				\end{equation*}
				Again there is no value of $f$ for which $z_1$ is real for any $\lambda>1$.\par
				To sum up, around $\re z = \pm 1$ the solution is the same as in the outer region, consistent only for $f$ fixed to one of the limit values $0$ or $1$.

		\section{Phase II with $\oR$ corrections}
		\label{app:phaseIIORcorrections}
			In this appendix we complete the discussion of Section \ref{sect:oRcorrectionsWL} with the analysis of phase II when $\oR$ corrections are included.\par
			Phase two arises in the region $\lambda_{-} < \lambda < \lambda_{+}$ and corresponds to $A$ comparable to $m$. Formally, this means $A-m \to 0$ in the decompactification limit. When $\oR$ corrections are taken into account, the requirement for phase II is:
			\begin{equation}
				\left( A - m \right)R = \log \phi ,
			\end{equation}
			for some $\phi$ depending on the parameters $\lambda$ and $\chi$. Continuity of $A$ as a function of $\lambda$ imposes:
			\begin{equation*}
				\phi \xrightarrow{ \lambda \to \lambda_{-} } 0 , \quad \phi \xrightarrow{ \lambda \to \lambda_{+} } + \infty .
			\end{equation*}
			We can evaluate $\phi$ in the decompactification limit using equation \eqref{eq:approxexprA}, which, at large $R$, gives:
			\begin{equation*}
				1 - \lambda \left( 1 - \chi \right) = \frac{\lambda \chi}{\left( 1 + \phi \right)^{1/2}} .
			\end{equation*}
			Therefore $\phi$ can be approximated at leading order by
			\begin{equation}
				\phi = \left( \frac{ \lambda \chi }{1 - \lambda \left( 1 - \chi \right) } \right) ^2 -1 , 
			\end{equation}
			as was already done in \cite{Barranco:2014}.\par
			If we want to evaluate the Wilson loop for phase II with $\oR$ corrections, we have to calculate $\phi$ at the corresponding order, and solve equation \eqref{eq:approxexprA} consistently to obtain $A$. The correct expression for $\phi$ at next-to-leading order is:
			\begin{equation}
			\begin{aligned}
				\phi + 1 & = \left[ \frac{ \lambda \chi }{1 - \lambda \left( 1 - \chi \right) + \frac{1}{m R} \log \left( \frac{1}{4} \left[ \left( \frac{ \lambda \chi }{1 - \lambda \left( 1 - \chi \right) } \right) ^2 - 1 \right] \right) } \right]^2 \\
					& \approx \left( \frac{ \lambda \chi }{1 - \lambda \left( 1 - \chi \right) } \right) ^2  \left[ 1 - \frac{2}{mR} \log \left( \frac{1}{4} \left[ \left( \frac{ \lambda \chi }{1 - \lambda \left( 1 - \chi \right) } \right) ^2 - 1 \right] \right)  \right] .
			\end{aligned}
			\end{equation}
			As a consequence, also the coefficient $\cII$ in the eigenvalue distribution is modified.\par
			Most important, those finite radius corrections slightly move $A$ away from $m$. Consequently, phase II resembles either phase I or III. More specifically, for values of $\lambda$ such that $\phi < 1$, then $A< m$ and the resonances fall out of the domain of $\rho $, so phase II is qualitatively analogous to phase I. Conversely if $\phi >1$ then $A>m$ and phase II behaves like phase III, in the sense that one should look for the saddle point in three different sub-regions for $z$.

		\section{Wilson loop derivatives}
		\label{app:WLfirstderiv}
			Here we write explicitly and phase by phase first order derivatives of the energy of Wilson loops $\frac{ \partial \mathcal{W}  }{\partial \lambda}$, where we recall that
			\begin{equation*}
				\mathcal{W} = - \frac{1}{RN} \log \langle W \left( f \right) \rangle .
			\end{equation*}
			Those expressions determine the order of each phase transition.
			
			\subsubsection{Two mass scales without FI term}
				The derivative of the Wilson loop energy in antisymmetric representation, for the case of masses $\pm m$ and without FI term is:
			\begin{equation}
			\begin{aligned}
				& \frac{ \partial \mathcal{W} ^{(0)} }{\partial \lambda}  \vert_{\mathrm{I}} = - m f \left( 1 - f \right) , \\
				& \frac{ \partial \mathcal{W} ^{(0)} }{\partial \lambda} \vert_{\mathrm{II}} = - m \left[ f \left( 1 - f \right) - \frac{ \lambda^2 - 1 }{4 \lambda^2 } \right] , \\
				& \frac{ \partial \mathcal{W} ^{(0)} }{\partial \lambda} \vert_{\mathrm{III}} = - m \left[ f \left( 1 - f \right) - \chi b (f) \right] ,
			\end{aligned}
			\end{equation}
			where to encode the derivatives of $\beta_{\lambda}$ in \eqref{eq:betashortcut}, we have introduced:
			\begin{equation}
			\label{eq:auxbderivfunction}
			 b (f) := \begin{cases} f ,  \\  \frac{1}{2}\left( 1- \frac{ \chi }{2}\right) , \\  1 - f ,  \end{cases} \  \begin{matrix} 0 & \le f < & \frac{\lambda -1}{2 \lambda} - \frac{\chi}{2} , \\ \frac{\lambda -1}{2 \lambda} & < f < &  \frac{\lambda +1}{2 \lambda} , \\  \frac{\lambda +1}{2 \lambda} + \frac{\chi}{2} & < f \le & 1 . \end{matrix}
			\end{equation}
			
			\subsubsection{Two mass scales with FI term}
				The antisymmetric Wilson loop in presence of two mass scales, each one associated to a set of $N_f$ hypermultiplets, was given is Subsection \ref{sec:FIWL2fameval}. We take first order derivatives of $\mathcal{W}$ with respect to $\lambda$, in the decompactification limit. Those are given in each phase by:
				\begin{equation}
				\label{eq:WL1d2ms}
				\begin{aligned}
					& \frac{ \partial \mathcal{W} ^{(0)} }{\partial \lambda} \vert_{\mathrm{I}} = - m f \left( 1 - f \right) , \\
					& \frac{ \partial \mathcal{W} ^{(0)} }{\partial \lambda} \vert_{\mathrm{II}} = - m  \left( \frac{ m_{-} }{ 2 m_{-} - m \lambda } \right)^3 \left[ f \left( 1 - f \right) \left( 6 - 5 \frac{m}{m_{-} } \lambda \right) + \frac{1}{4} \left(2 - 7 \frac{ m }{m_{-}} \lambda  + 6 \frac{ m^2 }{m_{-}^2} \lambda^2 - \frac{ m^3 }{m_{-}^3} \lambda^3 \right) \right]     , \\
					& \frac{ \partial \mathcal{W} ^{(0)} }{\partial \lambda} \vert_{\mathrm{IIIa}}= \begin{cases} - m \frac{1 - \chi }{\left( 1 - 2 \chi \right)^2} \left[ \left(1 - 3 \chi \right) f \left( 1 - f \right) - \frac{ 1 - 2 \chi}{4 \left( 1 - \chi \right)} \right] , \\ - m \frac{1 - \chi }{\left( 1 - 2 \chi \right)^2} \left[  \left(1 - 3 \chi^2 \right) f \left( 1 - f \right) + \frac{1}{2} \left( 1 - 2 \chi \right)\left( 1 + \chi \right)^2 + \frac{\left( 1 - 2 \chi \right)}{4 \left( 1 - \chi \right)}\left( 1 - 2 \chi \frac{m_{-}}{m \lambda} \right) \right] , \end{cases} \\
					& \frac{ \partial \mathcal{W} ^{(0)} }{\partial \lambda} \vert_{\mathrm{IIIb}}= - m \left( \frac{ m_{-}}{2 m_{-} - m_{+}} \right)^2  \left[ f \left( 1 - f \right)\left(3 - \frac{2 m_{+}}{m_{-}} \right) - \frac{1}{4} \left( 3 - \frac{2 m_{+}}{m_{-}} \right) + \lambda^{-2} \left( \frac{ m_{+} ^2 }{m^2} \right) \left( 1 - \frac{ m_{+} }{2m_{-}} \right)^2 \right] , \\
					& \begin{aligned} \frac{ \partial \mathcal{W} ^{(0)} }{\partial \lambda} \vert_{\mathrm{IV}} = - m \left( \frac{ m \lambda \left( 1 - \chi \right) }{ 2 m \lambda \left( 1 - \chi \right) - m_{+} } \right)^3 & \left[  f \left( 1 - f \right) \left( 6 - 9 \frac{m_{+}}{m \lambda \left(1 - \chi \right)} + 4 \left( \frac{m_{+}}{m \lambda \left(1 - \chi \right)} \right)^2 \right) \right. \\ & \left. + \left( 2 - \frac{m_{+}}{m \lambda \left( 1 - \chi \right)} \right) \frac{ \partial \ }{\partial \lambda} \left( \lambda \psi_{\lambda} \right) - 2 \frac{m_{+}}{m \lambda \left( 1 - \chi \right)} \psi_{\lambda}  \right] , \end{aligned} \\
					& \frac{ \partial \mathcal{W} ^{(0)} }{\partial \lambda} \vert_{\mathrm{V}} = - m \left[ f \left( 1  - f \right) - \chi b (f) \right]   ,
				\end{aligned}
				\end{equation}
				where for phase V the function $b (f)$ is analogous to the function defined in \eqref{eq:auxbderivfunction}, up to an obvious modification of the discontinuity points according to the insertion of FI coupling. It is not hard to check that this expression is continuous at first and last critical values, as long as solution (ii) holds for phases IV and V. However, at the second and third critical values, $\frac{ \partial \mathcal{W} ^{(0)} }{\partial \lambda}$ fails to be continuous. This means that transitions from phase II to III and from phase III to IV are of first order. Taking higher derivatives one arrives to the conclusion that the transition from phase I to II and from phase IV to V are of second order.

			\subsubsection{One mass scale with FI term}
				The same calculations are worked out when only one set of hypermultiplets appears. The antisymmetric Wilson loop was obtained in Subsection \ref{sec:FIWL1fameval}. Passing to the logarithm and taking the derivative with respect to the control parameter $\lambda$, we get:
				\begin{equation}
				\label{eq:WL1d1ms}
				\begin{aligned}
					& \frac{ \partial \mathcal{W} ^{(0)} }{\partial \lambda}   \vert_{\mathrm{I}} = - m \frac{ \left( 1 + \frac{5 \chi}{2} \right) \left( 1 + \frac{\chi}{2} \right) }{ \left( 1 + \frac{3 \chi}{2} \right)^2 } \left[ f \left( 1 - \frac{\chi}{2} - f \right)  + \frac{\chi^2 \left( 1 - \frac{\chi}{2} \right)^2}{ 4 \left( 1 + \frac{5 \chi}{2} \right) \left( 1 + \frac{\chi}{2} \right) } \right]   , \\
					& \begin{aligned}  \frac{ \partial \mathcal{W} ^{(0)} }{\partial \lambda}   \vert_{\mathrm{II}} =  - m & \left( \frac{ m_{-}}{ 2 m_{-} - m \lambda \left( 1 - \frac{\chi}{2} \right) } \right)^3  \left[ f \left( 1 - \frac{\chi}{2} - f \right) \left( 6 - 5 \frac{ m \lambda \left( 1 - \frac{\chi}{2} \right) }{m_{-}} \right) \right. \\ & \left. +  \left( 1 - \frac{\chi}{2} \right)^2 \left( 2 - 7 \frac{ m \lambda \left( 1 - \frac{\chi}{2} \right) }{m_{-} } + 6 \left( \frac{ m \lambda \left( 1 - \frac{\chi}{2} \right) }{m_{-} } \right)^2 -  \left( \frac{ m \lambda \left( 1 - \frac{\chi}{2} \right)}{m_{-} } \right)^3  \right) \right] , \end{aligned} \\
					& \frac{ \partial \mathcal{W} ^{(0)} }{\partial \lambda}   \vert_{\mathrm{III}} = \begin{cases} - m  f \left( 1 - \frac{\chi}{2} - f \right)  , \\ - m  \left( f - \frac{\chi}{2} \right) \left( 1  - f \right)  . \end{cases}
				\end{aligned}
				\end{equation}
				This expression is discontinuous at first critical value $\lambda = \frac{m_{-}}{m \left( 1 + \frac{\chi}{2} \right)}$ and continuous at second critical value $\lambda = \frac{m_{-}}{m \left( 1 - \frac{\chi}{2} \right)}$. Taking second order derivatives one concludes that the transition from phase I to II is first order and from phase II to III is second order.

	\end{appendix}

	\bibliographystyle{JHEP}
	\bibliography{sCS}
%	\printbibliography
	
\end{document}